\documentclass[sigplan,screen]{acmart}
\newcommand{\BSR}{\texttt{BSR}}
\newcommand{\SR}{\texttt{SR}}
\newcommand{\RACE}{\texttt{R2H}}
\newcommand{\ABFTOC}{\texttt{ABFT-OC}}

\usepackage{times}
\usepackage{graphicx}
\usepackage{xspace,shortcut}
\usepackage{comment}
\usepackage{color,soul}
\usepackage{tcolorbox}
\tcbuselibrary{skins}
\usepackage{listings}
\usepackage{amsmath}
\usepackage{multirow}
\usepackage{subcaption}
\usepackage{algcompatible}
\usepackage[ruled,linesnumbered]{algorithm2e}
\usepackage{graphics}
\def\BibTeX{{\rm B\kern-.05em{\sc i\kern-.025em b}\kern-.08em
    T\kern-.1667em\lower.7ex\hbox{E}\kern-.125emX}}

\definecolor{backcolour}{rgb}{0.95,0.95,0.92}

\usepackage{colortbl}
\usepackage{tikz}
\usetikzlibrary{positioning}
\newcommand*\Circled[1]{
	\tikz[baseline=(char.base)]{\node[
        shape=circle, draw=none,  thick, 
        fill=gray!40,inner sep=0.6pt] (char) 
    {\textcolor{black}{\sffamily#1}}; 
}}

\newtcbox{\inlinebox}[1][]{enhanced,
 box align=base,
 nobeforeafter,
 colback=gray!40,
 colframe=white,
 size=small,
 left=0pt,
 right=0pt,
 boxsep=2pt,
 #1}

\copyrightyear{2023} 
\acmYear{2023} 
\setcopyright{acmcopyright}\acmConference[PPoPP '23]{The 28th ACM SIGPLAN Annual Symposium on Principles and Practice of Parallel Programming}{February 25-March 1, 2023}{Montreal, QC, Canada}
\acmBooktitle{The 28th ACM SIGPLAN Annual Symposium on Principles and Practice of Parallel Programming (PPoPP '23), February 25-March 1, 2023, Montreal, QC, Canada}
\acmPrice{15.00}
\acmDOI{10.1145/3572848.3577496}
\acmISBN{979-8-4007-0015-6/23/02}

\begin{document}

\title{Improving Energy Saving of One-sided Matrix Decompositions on CPU-GPU Heterogeneous Systems}


\author{Jieyang Chen}
\email{jchen3@uab.edu}
\affiliation{
  \institution{University of Alabama at Birmingham}
  \city{Birmingham}
  \state{Alabama}
  \country{USA}
}

\author{Xin Liang}
\email{xliang@uky.edu}
\affiliation{
  \institution{University of Kentucky}
  \city{Lexington}
  \state{Kentucky}
  \country{USA}
}

\author{Kai Zhao}
\email{kzhao@uab.edu}
\affiliation{
  \institution{University of Alabama at Birmingham}
  \city{Birmingham}
  \state{Alabama}
  \country{USA}
}

\author{Hadi Zamani Sabzi}
\email{hzama001@ucr.edu}
\affiliation{
  \institution{University of California, Riverside}
  \city{Riverside}
  \state{California}
  \country{USA}
}

\author{Laxmi Bhuyan}
\email{bhuyan@cs.ucr.edu}
\affiliation{
  \institution{University of California, Riverside}
  \city{Riverside}
  \state{California}
  \country{USA}
}

\author{Zizhong Chen}
\email{chen@cs.ucr.edu}
\affiliation{
  \institution{University of California, Riverside}
  \city{Riverside}
  \state{California}
  \country{USA}
}





\begin{abstract}
One-sided dense matrix decompositions (e.g., Cholesky, LU, and QR) are the key components in scientific computing in many different fields.
Although their design has been highly optimized for modern processors, they still consume a considerable amount of energy. 
As CPU-GPU heterogeneous systems are commonly used for matrix decompositions, in this work, we aim to further improve the energy saving of one-sided matrix decompositions on CPU-GPU heterogeneous systems.
We first build an Algorithm-Based Fault Tolerance protected overclocking technique (\ABFTOC) to enable us to exploit reliable overclocking for key matrix decomposition operations. 
Then, we design an energy-saving matrix decomposition framework, Bi-directional Slack Reclamation (\BSR), that can intelligently combine the capability provided by \ABFTOC~and DVFS to maximize energy saving and maintain performance and reliability.
Experiments show that \BSR~is able to save up to 11.7\% more energy compared with the current best energy saving optimization approach with no performance degradation and up to 14.1\% $Energy \times Delay^2$ reduction.
Also, \BSR~enables the Pareto efficient performance-energy trade-off, which is able to provide up to 1.43$\times$ performance improvement without costing extra energy.
\end{abstract}



\maketitle


\section{Introduction}
To meet performance requirements for current mission-critical scientific computing, millions of computing cores are equipped in modern High Performance Computing (HPC) systems consuming tens of megawatts of power~\cite{top500}.
With the increasing need for higher performance, it is anticipated that future HPC systems will consist of even more computing cores and consume more power.
As HPC systems are achieving higher parallelism, how to achieve high performance and energy efficiency while ensuring computing reliability has become a critical challenge for scientific computing.

As the type of processor that contributes the most of the computing parallelism in many current and future HPC systems, Graphics Processing Units (GPUs), equipped with thousands of low-power cores, offer high computational power and energy efficiency.
Many applications and libraries have been designed and optimized for GPU accelerators~\cite{burau2010picongpu, ku2009full,chang2004numerical, chen2019tsm2, rivera2021tsm2x, tian2021revisiting, chen2021accelerating, tian2020cusz, moreland2016vtk, chetlur2014cudnn}.
Benefiting from the fact that GPUs are designed for highly parallelizable computations while CPUs are more efficient with serial computations, CPUs and GPUs that are linked through fast interconnections ~\cite{li2019evaluating, li2018tartan} are usually used together to form heterogeneous systems that can efficiently handle a large spectrum of scientific computing workloads.
Many scientific applications or software begin to have an optimized design for CPU-GPU heterogeneous systems such as the MAGMA linear algebra library~\cite{dghklty14}.

One-sided dense matrix decomposition such as Cholesky, LU, and QR play a pivotal role in many scientific applications. 
Their state-of-the-art designs for CPU-GPU heterogeneous systems are proposed in \cite{tdb10,tnld10}, and they have been highly optimized in the MAGMA library and used as key computational kernels by many applications across different fields~\cite{zegard2013toward, jalili2012large, ukidave2016mystic, guyon2017adaptive, murray2012collapsed, he2013multigrid, joubert2018attacking}.

Many works have been done to improve the energy efficiency of linear algebra operations.
\cite{chen2016greenla} proposed to use a DVFS-based approach to optimize matrix decompositions.
\cite{zamani2019greenmm, zamani2020saou, cavelan2015voltage} seek to use reduced core supply voltage to reduce the energy consumption of matrix-matrix multiplication operations.
Although much work has been done to improve the energy saving of matrix decomposition on CPU-GPU heterogeneous systems, 
it is still desirable to further improve their energy saving since matrix decompositions as they still consume a considerable amount of energy. Improving the energy saving of matrix decomposition can lead to more energy-efficient scientific computing. 
However, the major challenge, as pointed out in~\cite{zamani2019greenmm, zamani2020saou, tan2011analyzing, leng2015safe, leng2016guardband}, is that aggressive energy-saving optimizations can weaken the reliability of the system and cause performance degradation, which is unacceptable for time-sensitive and mission-critical scientific applications. 

In this work, we aim to further improve the energy saving of one-sided matrix decompositions on CPU-GPU heterogeneous systems while maintaining performance and reliability.
We first build an Algorithm-Based Fault Tolerance protected overclocking technique (\ABFTOC) to enable us to exploit reliable overclocking for key matrix decomposition operations. 
Then, we design an energy-saving matrix decomposition framework, Bi-directional Slack Reclamation (\BSR), that can intelligently combine the capability provided by \ABFTOC~and Dynamic Voltage and Frequency Scaling (DVFS) to maximize energy saving and maintain performance and reliability.
Also, \BSR~enables the Pareto efficient performance-energy trade-off.
Specifically, our contributions are listed as follows:
\begin{itemize}
    \item We propose the first adaptive algorithm-based fault tolerance protected overclocking technique (\ABFTOC) for matrix decompositions on CPU-GPU heterogeneous systems. Overclocking with an optimized voltage guardband can enable us to exploit higher clock frequencies with higher energy efficiency. However, aggressive overclocking can decrease system reliability, so we propose to couple ABFT with overclocking to enable trustable computation. To reduce fault tolerance overhead, we further propose a lightweight adaptive-ABFT technique that automatically adjusts its fault tolerance strength according to the error rate.
    \item Next, based on \ABFTOC, we propose a novel slack-based energy saving framework - Bi-directional Slack Reclamation (\BSR), which aims to exploit \textit{slack}, processor idle time, to save energy and enable flexible Pareto efficient performance-energy trade-off. Different from existing works, \BSR~reclaims slack in both directions using both \ABFTOC~and DVFS to save more energy and enable performance improvement.
    \item We implement our \BSR~on three key one-sided matrix decompositions: Cholesky, LU, and QR. We evaluate our implementation on a modern CPU-GPU heterogeneous system with Nvidia GPU. Experiments show that \BSR~is able to save up to 11.7\% more energy compared with the current best energy saving optimization approach with no performance degradation and up to 14.1\% $Energy \times Delay^2$ reduction. Also, \BSR~enables the Pareto efficient performance-energy trade-off, which is able to provide up to 1.43$\times$ performance improvement without costing extra energy.

\end{itemize}

\begin{figure}[ht]
    \centering
    \begin{subfigure}[t]{0.25\textwidth}
    \includegraphics[width=\textwidth]{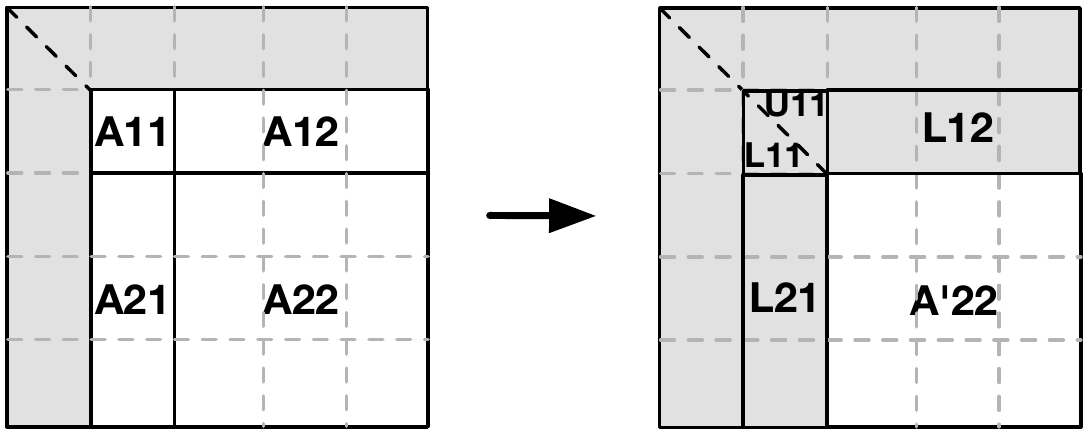}
    \vspace*{-0.5em}
    \caption{Blocked algorithm}
    \end{subfigure}
    \begin{subfigure}[t]{0.45\textwidth}
    \includegraphics[width=\textwidth]{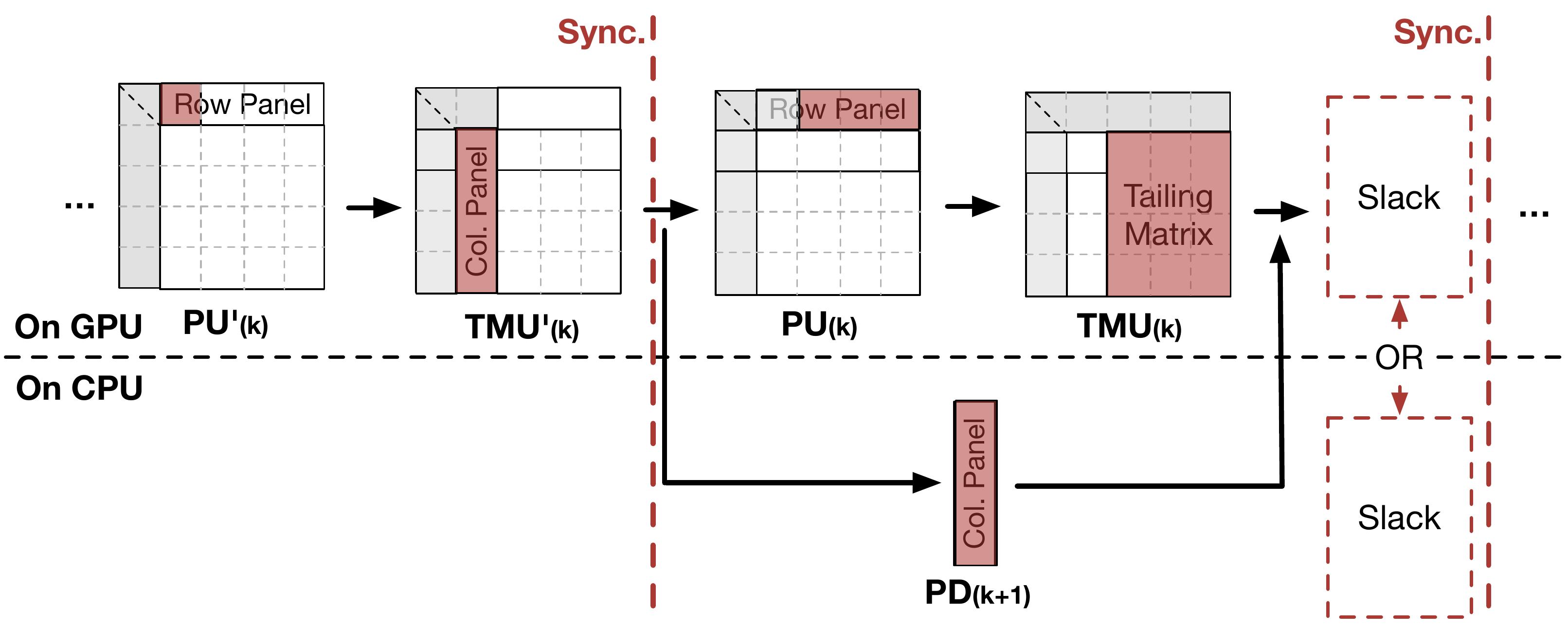}
    \vspace*{-1em}
    \caption{State-of-the-art design on CPU-GPU heterogeneous systems
    }
    \end{subfigure}
    \vspace*{-1em}
    \caption{One iteration of LU decomposition}
    \label{example-lu}
    \vspace*{-2em}
\end{figure}

\begin{figure}[ht]
    \centering
    \begin{subfigure}[t]{0.23\textwidth}
    \includegraphics[width=\textwidth]{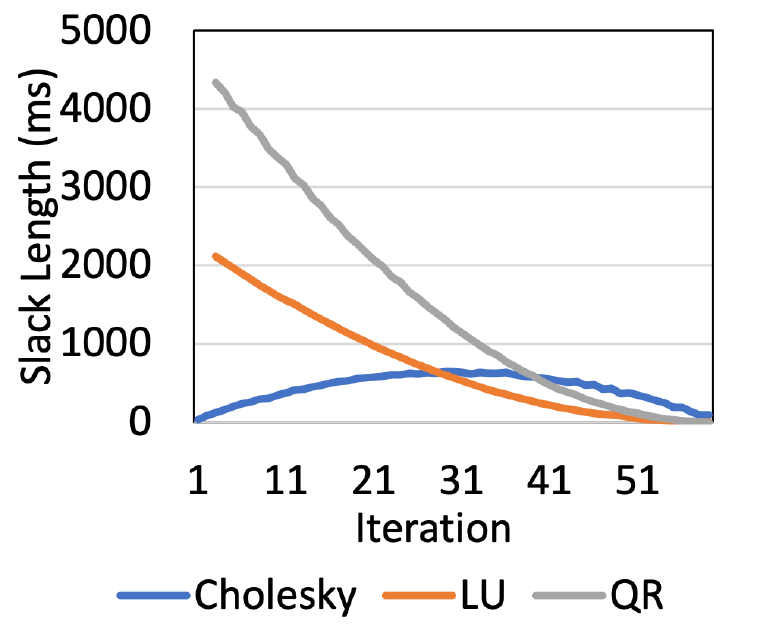}
    \caption{Double precision}
    \end{subfigure}
    \begin{subfigure}[t]{0.23\textwidth}
    \includegraphics[width=\textwidth]{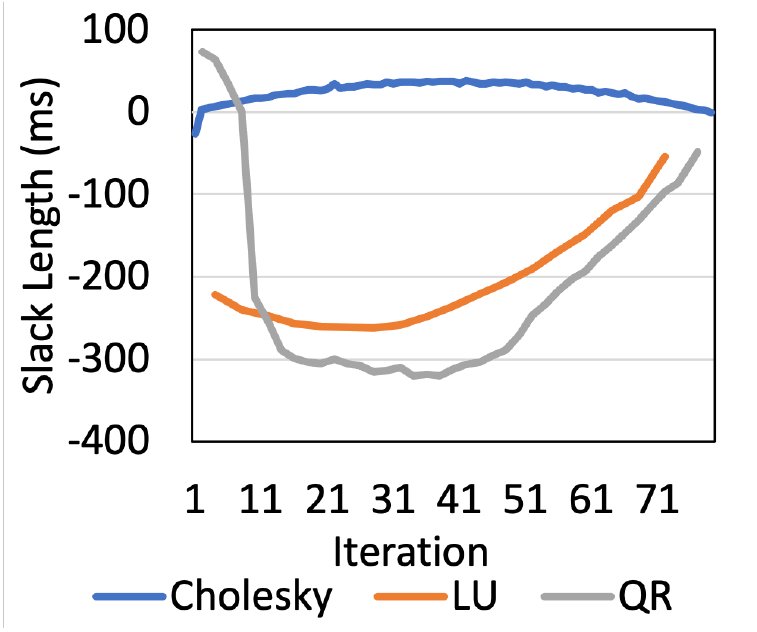}
    \caption{Single precision}
    \end{subfigure}
    \vspace*{-0.5em}
    \caption{Slacks occur when decomposing a $30720 \times 30720$ matrix on our heterogeneous system. Block size is optimized for performance.
    Positive values represent slacks on the CPU side and negative values represent slacks on the GPU side.}
    \label{slack}
    \vspace{-1em}
\end{figure}

\section{Related Works and Problem Statement}
\label{problem}
In this section, we first introduce the design of state-of-the-art matrix decomposition on CPU-GPU heterogeneous systems and we focus on discussing their key computing characteristics.
Then, we review how existing works leverage such computing characteristics to optimize for energy efficiency.
Finally, we formulate our research problem and challenges.

\subsection{State-of-the-art matrix decompositions}
The state-of-the-art matrix decompositions for CPU-GPU heterogeneous systems use the blocked version matrix decomposition algorithms.
Blocks, logically divided sub-matrices, form \textit{Panel} and \textit{Trailing Matrix}.
The decomposition process begins from the up left corner of the matrix and moves towards the down right corner iteratively.
An illustration of one iteration of the LU decomposition is shown in \textbf{Figure~\ref{example-lu}(a)}.
Each iteration includes three major operations:  \Circled{1} \textit{Panel decomposition} (PD): $L\cdot1 \times U11 \leftarrow A\cdot1$; \Circled{2} \textit{Panel update} (PU): $ U12 \leftarrow (L11)^{-1} \times A12$; and \Circled{3} \textit{Trailing matrix update} (TMU): $A'22 \leftarrow A22 - L21 \times U12$. 
Cholesky, LU, and QR decomposition all share similar three operations.
On CPU-GPU heterogeneous systems, the three operations are assigned to different processors based on their characteristics. 
PD is assigned to the CPUs since it is highly sequential.
PU and TMU are assigned to the GPUs as they are high parallelizable. 
As illustrated in \textbf{Figure~\ref{example-lu}(b)}, to overlap the computation on CPUs and GPUs, a look-ahead optimization~\cite{kurzak2006implementing} is used that allows the partial PU and TMU to be done first (i.e., PU' and TMU'), so that the PD of the next iteration can be done with the rest of PU and TMU concurrently.
Depending on the computational power of the CPU/GPU and the amount of workload assigned during decomposition, those concurrent tasks may finish at different times, which leads to idle computing cycles on the CPU or GPU. 
The idle is called \textit{slack}. 
\textbf{Figure \ref{slack}} show how slack length can change during Cholesky, LU, and QR decompositions on our test platform.

\subsection{Existing slack-based energy saving}
Matrix decompositions have been designed to maximize their usage on highly optimized BLAS-3 GPU kernels, so their energy efficiency is inherently high, which leaves limited room for further optimization.
As for now, the most effective class of energy-saving optimizations for matrix decompositions on CPU-GPU heterogeneous systems is DVFS-based approaches, which aim to exploit different energy-saving techniques when there are slacks.

\begin{figure}[h]
    \vspace*{-0.75em}
    \centering
    \includegraphics[width=0.45\textwidth]{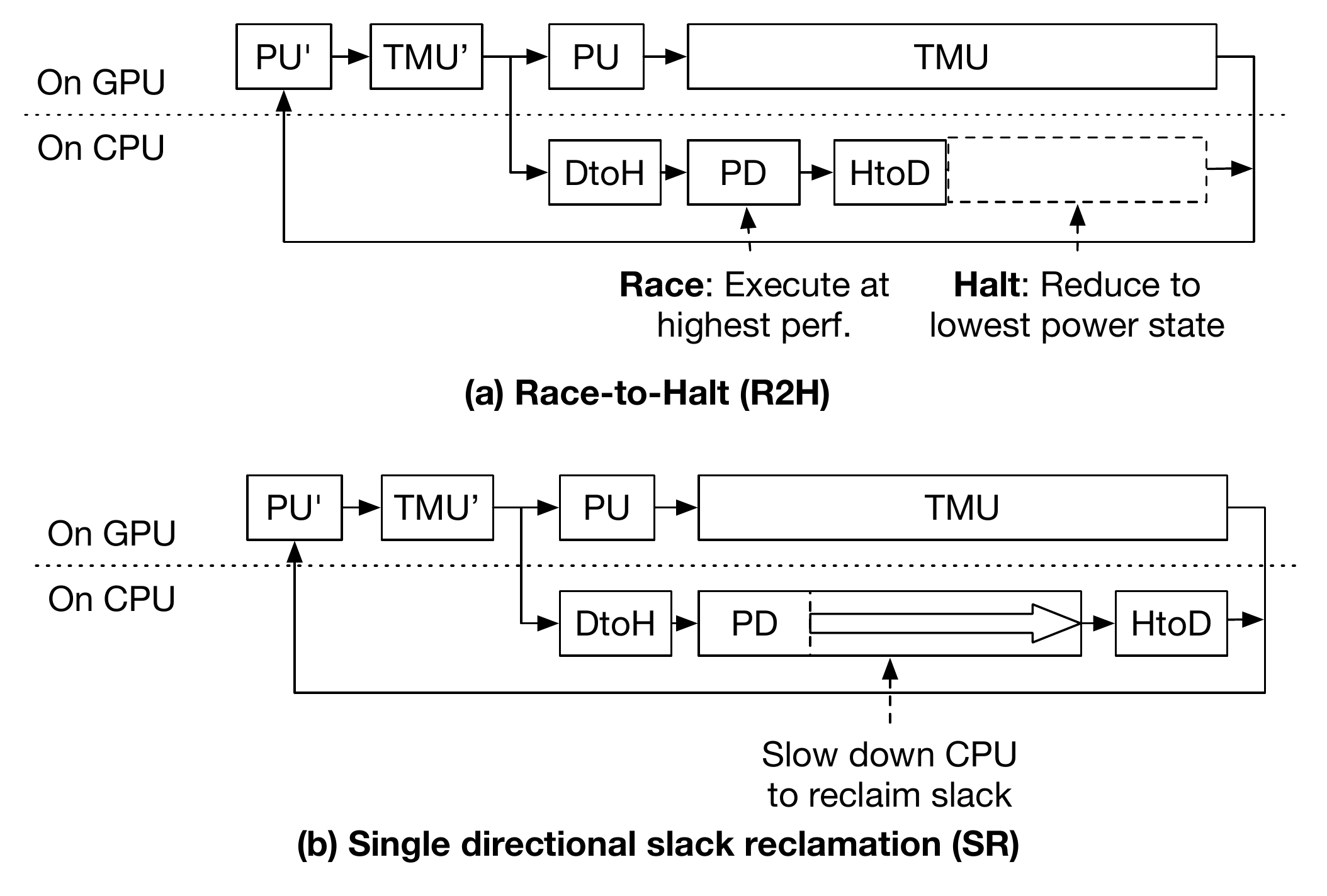}
    \vspace*{-1em}
    \caption{Existing slack-based energy saving}
    \label{r2h-sr}
    \vspace*{-1.5em}
\end{figure}

There are two main strategies for optimizing energy costs: \textit{Race-to-Halt (\RACE)}~\cite{ishihara1998voltage, rizvandi2011some, tan2015slow} and \textit{Slack Reclamation (\SR)}~\cite{chen2016greenla}.
As shown in \textbf{Figure~\ref{r2h-sr}}, the main idea of R2H is to timely reduce clock frequency to the minimum as soon as the tasks on the non-critical path finish.
The processor maintains its minimum clock frequency during the slack to save energy.
This strategy is usually implemented by the hardware or the operating system leveraging their workload monitoring capabilities.
\SR~saves energy by slowing down the tasks on the non-critical path. 
The reason this strategy can save energy is due to the relation between the dynamic power of the processor and its clock frequency $P_{dynamic} \propto f^{2.4}$ \cite{cal14}.
Theoretically, \SR~is able to save more energy compared with \RACE~\cite{chen2016greenla}.
Since the processor's clock frequency need to be adjusted before the execution of each task and the length of slack can change as shown in \textbf{Figure \ref{slack}}, some form of computation pattern prediction is necessary.
In the start-of-the-art \SR~\cite{chen2016greenla}, the authors propose to predict computation patterns leveraging algorithmic knowledge in matrix decompositions.

\subsection{Motivation of further improving energy saving}
Despite a lot of research efforts have been made to improve the energy saving of matrix decomposition on CPU-GPU heterogeneous systems, it is still desirable to further improve their energy saving since matrix decompositions are heavily used in many scientific applications. 
Thus improving the energy saving of matrix decomposition can lead to more energy-efficient scientific computing.

\subsection{Challenges of further improving energy saving}
\subsubsection{Performance degradation} DVFS is designed to enable performance-energy trade-off while maintaining processor reliability. 
So, existing DVFS-based energy-saving techniques can only be applied to tasks on the non-critical path to avoid negatively impacting the overall performance.
This has already been extensively exploited by existing works.
To save even more energy, the only other choice is to apply DVFS-based energy-saving techniques to tasks on the critical path, however, this will inevitably lead to performance degradation to the overall decomposition since modern CPU and GPU processors tend to have better energy efficiency when running at lower clock frequencies. 

\subsubsection{Reliability degradation} Other approaches such as processor undervolting can also be used to reduce the energy cost of computation.
Since it works by decreasing the core supply voltage while maintaining the same clock frequencies, it can save energy without performance degradation.
However, they can decrease system reliability~\cite{zamani2019greenmm, zamani2020saou, cavelan2015voltage}. 
Such reliability degradation can be manifested as hard errors (e.g., process or system crash) or SDCs (e.g., incorrect calculation, bit-flips in memory cells), which can seriously decrease the reliability of matrix decomposition.
Although ABFT has been used with undervolting in~\cite{zamani2019greenmm, cavelan2015voltage} to improve the energy efficiency of matrix-matrix multiplications and ensure computing correctness, applying existing ABFT techniques can still bring considerable performance overhead. 
This overhead can be especially high for matrix decompositions since the iterative computing fashion is prone to error propagation, which needs the strongest variant of ABFT, \textit{full checksum ABFT}~\cite{chen2018fault}, to provide sufficient protection.

\subsection{Research questions}
In this work, we try to answer the following research questions: 



\noindent
\inlinebox{RQ:1} How to further improve energy saving of matrix decompositions on CPU-GPU heterogeneous system beyond the state-of-the-art works?

\noindent
\inlinebox{RQ:2} How to maximize energy saving for matrix decomposition while maintaining both performance and reliability at the same time?

\noindent
\inlinebox{RQ:3} How to enable performance-energy trade-off in matrix decomposition?


\begin{figure}[!ht]
    \centering
    \includegraphics[width=0.4\textwidth]{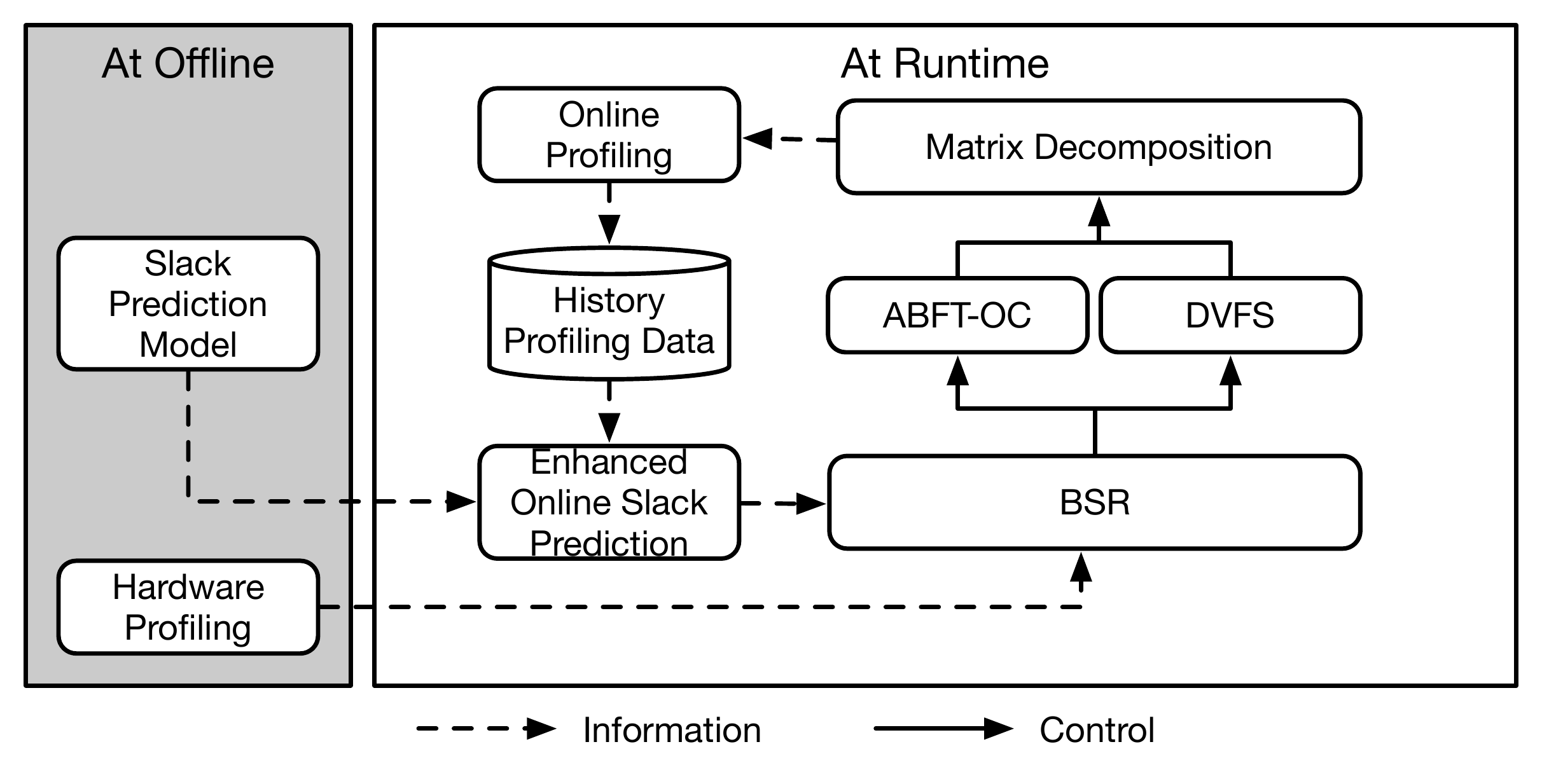}
    \vspace*{-1em}
    \caption{Overview of our energy-saving matrix decomposition framework}
    \label{overview}
    \vspace*{-2em}
\end{figure}
\section{Design of Energy-Saving Matrix Decomposition}
\label{bsr}
In this work, we propose to build a matrix decomposition framework that maximizes energy saving while maintaining both performance and reliability at the same time.
\textbf{Figure \ref{overview}} shows the overview of our framework.
We first focus on enabling reliable computation when overclocking by coupling ABFT with overclocking - \ABFTOC.
To reduce fault tolerance overhead, we further propose a lightweight adaptive-ABFT technique that automatically adjusts its fault tolerance strength according to the error rate. Next, based on \ABFTOC, we propose a novel slack-based energy saving optimization framework - \BSR, which aims to exploit slack, to save energy and enable flexible Pareto efficient performance-energy trade-off. 
Different from existing works, \BSR~reclaims slack in both directions using both \ABFTOC~and DVFS to save more energy and enable performance improvement.

\subsection{Adaptive Algorithm-Based Fault Tolerance Protected Overclocking (\ABFTOC)}
\label{beyong-dvfs}
\begin{figure}[h!]
	\begin{subfigure}[t]{0.4\textwidth}
    \includegraphics[width=\textwidth]{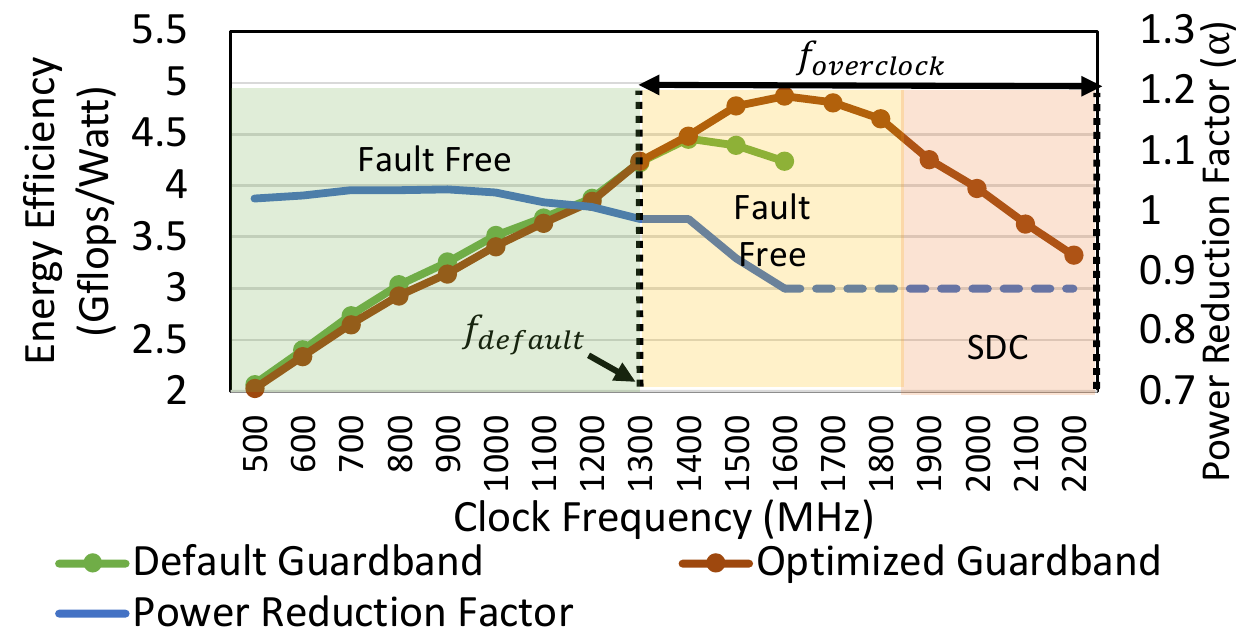}
    \caption{GPU energy efficiency}
    \end{subfigure}
    \begin{subfigure}[t]{0.35\textwidth}
    \includegraphics[width=\textwidth]{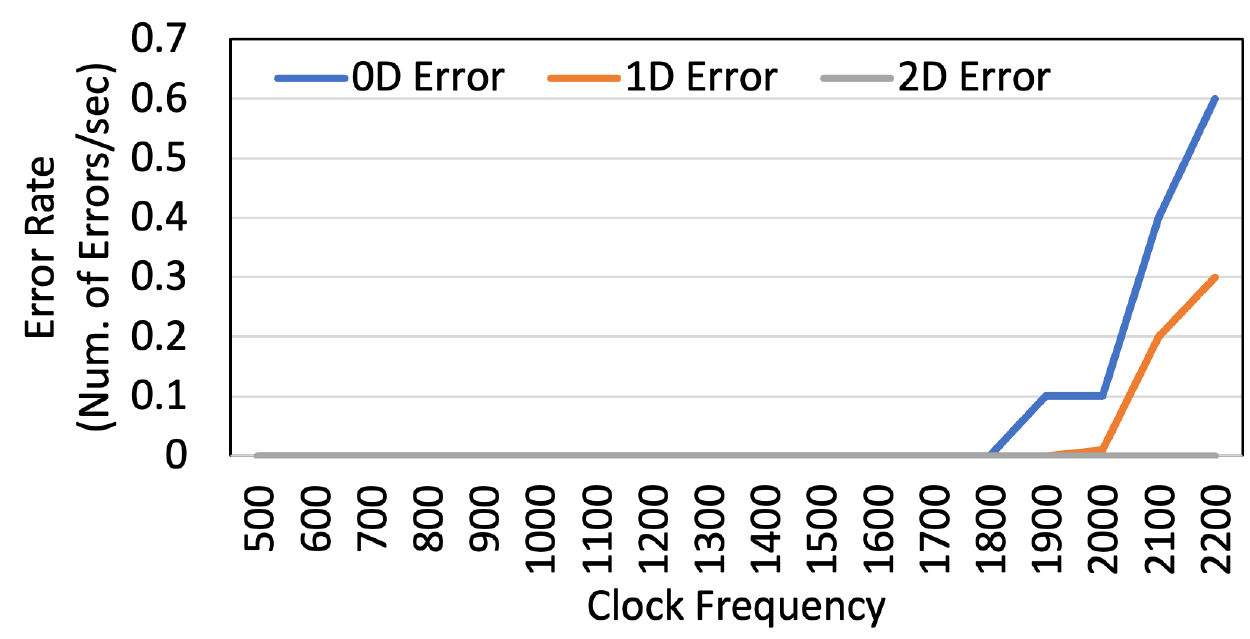}
    \caption{GPU SDC error rate}
    \end{subfigure}
    \begin{subfigure}[t]{0.4\textwidth}
    \includegraphics[width=\textwidth]{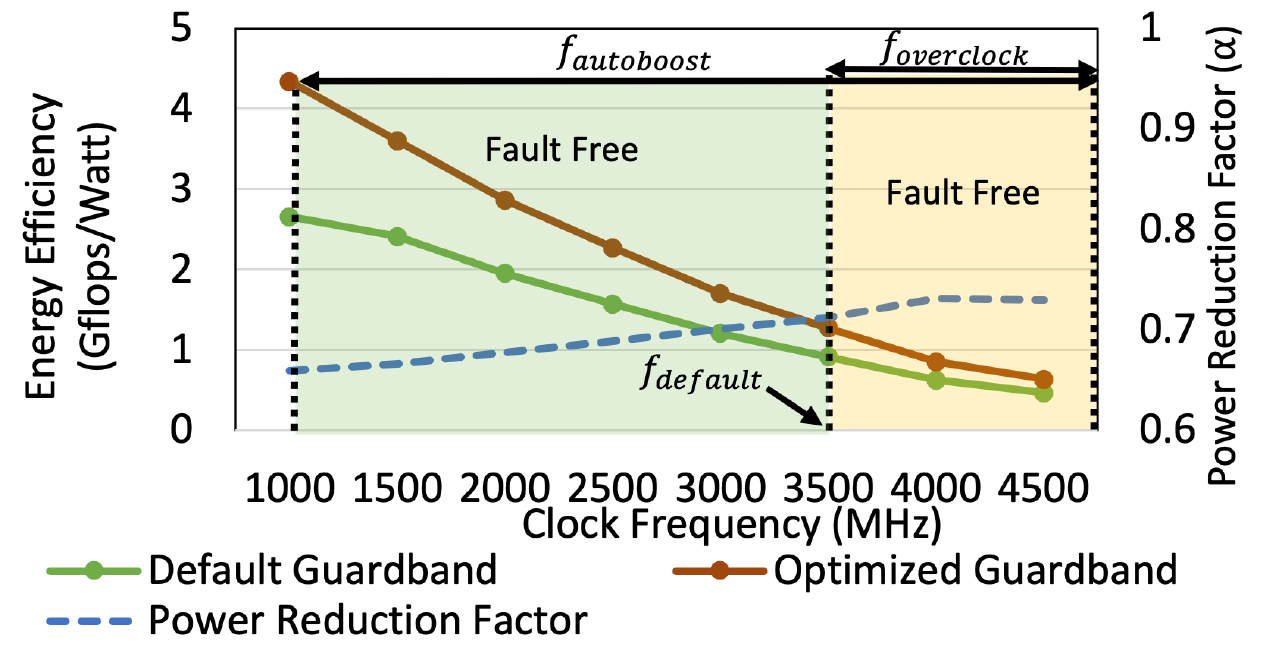}
    \caption{CPU energy efficiency}
    \end{subfigure}
    \begin{subfigure}[t]{0.2\textwidth}
    \includegraphics[width=\textwidth]{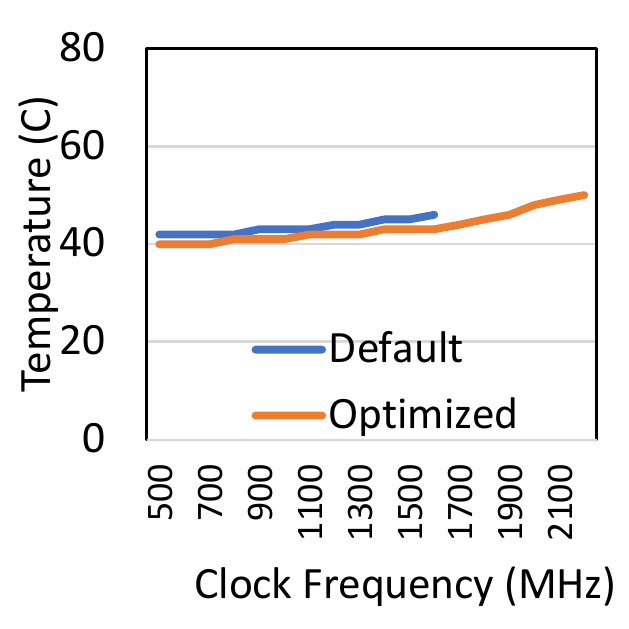}
    \caption{Maximum sustained GPU core temperature}
    \end{subfigure}
    \begin{subfigure}[t]{0.2\textwidth}
    \includegraphics[width=\textwidth]{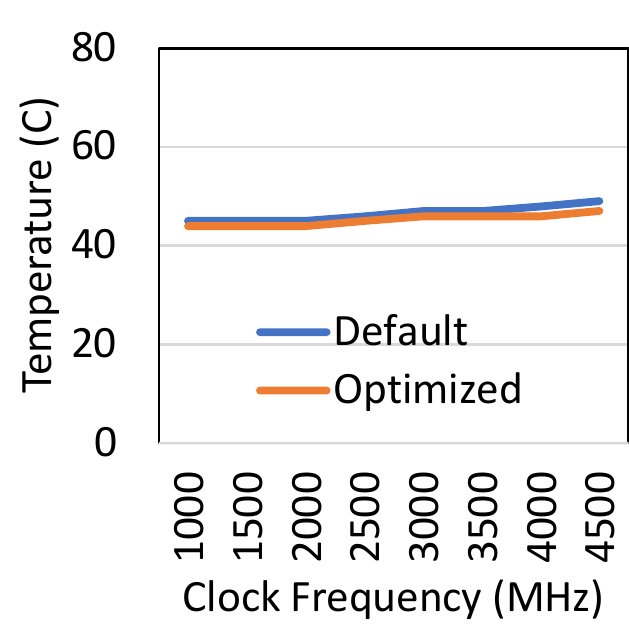}
    \caption{Maximum sustained CPU core temperature}
    \end{subfigure}
    \caption{Profiling results of our testing CPU and GPU 
    }
    \label{profile-overclock}
    \vspace*{-1em}
\end{figure}

To design a technique that maximize energy saving for matrix decompositions, we seek hardware energy optimization techniques beyond DVFS.
DVFS has been extensively used for energy saving by both hardware and applications.
It optimizes energy efficiency by lowering the core voltage ($V_{dd}$) with the decrease of clock frequency for reducing energy consumption.
However, lowering frequency can inevitably cause performance degradation.
Processor voltage guardband optimization largely mitigates this issue by allowing lowering of the core voltage without decreasing clock frequency or overclocking without violating the hardware power limit.

\subsubsection{Voltage guardband optimization for overclocking}
In this work, we define \textit{overclocking} as the processor state where it sustains at a higher-than-default clock frequencies.
\textbf{Figure \ref{profile-overclock}} (a) shows the achievable overclocking frequency range and their energy efficiency of our test GPU at different clock frequencies after we apply voltage guardband optimization.
Please note unlike previous works that were based on Windows-based GPU driver~\cite{zamani2019greenmm, leng2016guardband, leng2013gpuwattch} where the core voltage can be directly adjusted and monitored, the Linux-based GPU driver does not allow us to directly control and monitor the GPU core voltage. 
Even though we find that optimizing the voltage guardband of GPU on Linux is still achievable through the clock offset command of the NVML API on Linux-based GPU driver.
We omit the details due to the page limit.
CPU undervolting can be directly achieved on the Linux system. We set the offset of the CPU core voltage using a third-party tool \texttt{intel-undervolt}.
\textbf{Figure \ref{profile-overclock}} (c) shows the CPU energy efficiency before and after we set the optimized voltage guardband.
Please note unlike our testing GPU, our testing CPU can achieve overclocking with the default guardband, but an optimized guardband can help us achieve higher energy efficiency.

Finding the optimized guardband is done by gradually lowering specific power settings of CPU/GPU to the point where energy efficiency is maximized without process or OS crash. The whole process can be done in less than 30 minutes and it only needs to be done once during software installation. As optimized guardband can be workload-dependent, we specifically use the workload in matrix decomposition i.e., TMU on GPU and PD on CPU to find optimized guardband.
Also, as shown in \textbf{Figure \ref{profile-overclock}} (b), we observe that setting to extreme high clock frequencies for the GPU can weaken the reliability of computation e.g., SDCs.
So, we propose to incorporate fault tolerance with overclocking by designing~\ABFTOC.

\subsubsection{Design of \ABFTOC}
Reliable computation is the foundation of our optimized matrix decomposition.
As overclocking achieved through the use of optimized guardband can lead to SDCs, we propose to use ABFT~\cite{chen2016wu, wu2014ft, wu2017silent, chen2016magma, chen2016gpu, teresa2013ft,  yao2015detection, chen2018fault, liang2017correcting, chenextending, chen2016tao, li2019ft, hakkarinen2014fail, chen2009optimal, hakkarinen2012multilevel, tao2018improving, chen2008extending} to handle SDCs during matrix decompositions. 
Since the processor power state is under control and the corresponding SDC error rate is known, SDC error rate is predictable during matrix decompositions.
So we propose the first ABFT that can adjust its fault tolerance strength and overhead at runtime based on the predicted error rate to minimize fault tolerance overhead and ensure correctness.
SDC refers to the kind of error that only causes incorrect calculation results without process or system crash.
When using our optimized guardband, the SDC is caused by insufficient core voltage supply when at high clock frequencies. 
The rate of SDC can increase as we increase the clock frequency when we apply a optimized guardband at the same time.

\begin{figure}[!ht]
    \centering
    \includegraphics[width=0.4\textwidth]{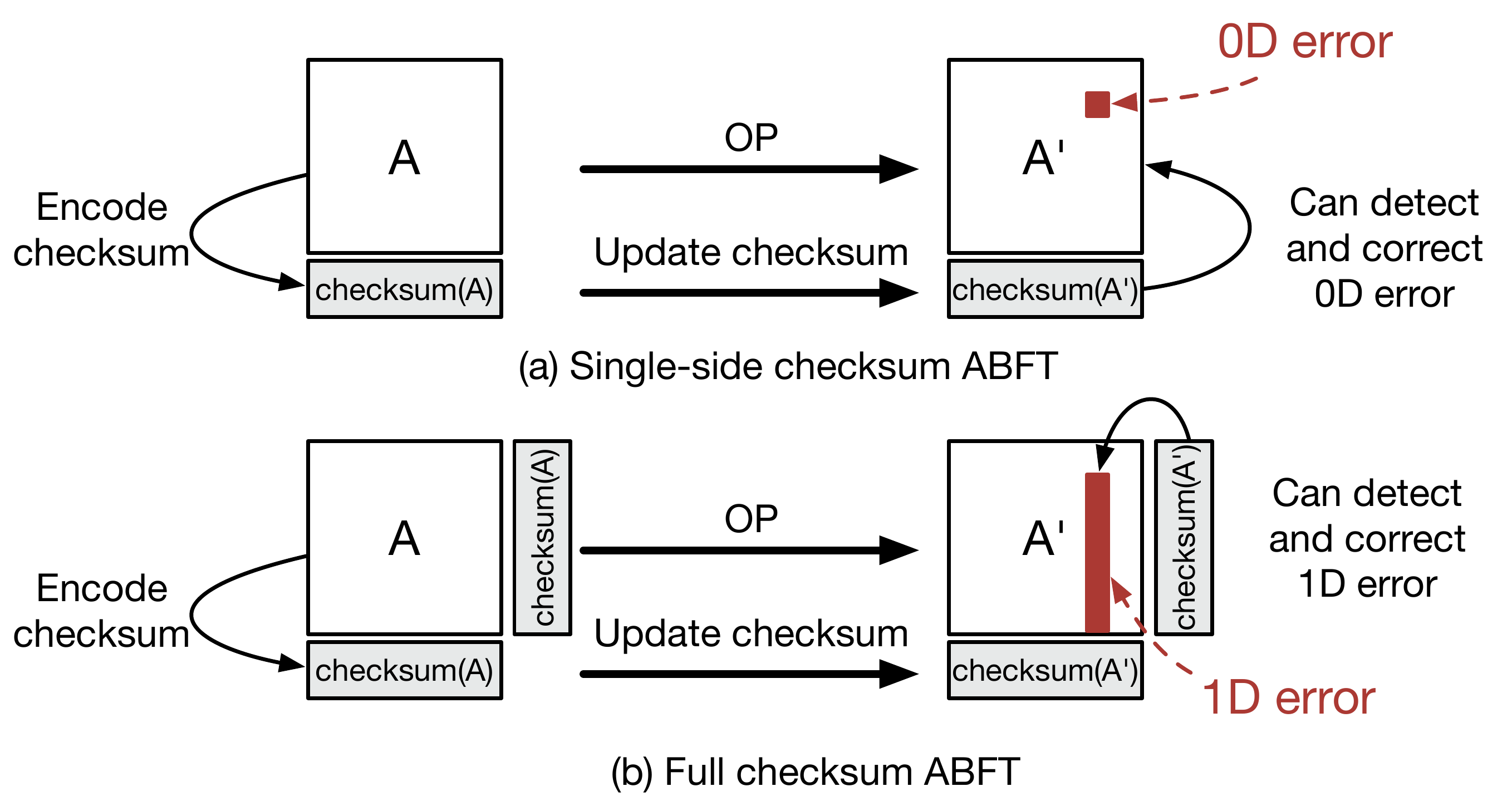}
    \vspace*{-0.5em}
    \caption{ABFT checksum for detecting and correcting SDCs in matrix operations}
    \label{single-full-abft}
    \vspace*{-1em}
\end{figure}

Depending on where the hardware fault occurs, it may be manifested as different kinds of SDC. 
For example, calculation error is usually caused by faults in the logic part of ALU or FPU. 
Memory storage error is usually caused by faults (e.g., bit flips) in the storage cells of DRAM, cache, or registers. 
For matrix operations, matrix elements can be repeatedly accessed to obtain final results. 
If an element whose value is corrupted gets repeatedly referenced, it may cause error propagation.
Depending on the cause of the error and the computation pattern (i.e., how data is used/reused) of a matrix operation, the error pattern can be different. 
The degrees of error propagation~\cite{chen2018fault} can be classified as: 0D, 1D, and 2D. \textbf{0D}: a single standalone error with no error propagation; \textbf{1D}: an error propagates to entire/part of one row/column; \textbf{2D}: an error propagates beyond one row/column.
So, we distinguish different degrees of error propagation in \textbf{Figure~\ref{profile-overclock}}.

\begin{table}[!ht]
\caption{Theoretical estimation on ABFT fault coverage (FC) on the TMU operation of the $5^{th}$, $10^{th}$, and $15^{th}$ iteration of LU decomposition if we apply different clock frequencies.}
\label{clock-abft}
\resizebox{1\columnwidth}{!}{%
\begin{tabular}{|c|c|c|c|c|c|c|}
\hline
\multicolumn{1}{|l|}{Iter.} & ABFT & 1800MHz & 1900MHz & 2000MHz & 2100MHz & 2200MHz \\ \hline
\multirow{2}{*}{$5^{th}$} & Single & Fault-free & Full Coverage & 99.86\% & 97.51\% & 96.45\% \\   & Full & Fault-free & Full Coverage & Full Coverage & Full Coverage & Full Coverage \\ \hline
 \multirow{2}{*}{$10^{th}$} & Single & Fault-free & Full Coverage & 99.94\% & 98.92\% & 98.46\% \\   & Full & Fault-free & Full Coverage & Full Coverage & Full Coverage & Full Coverage \\ \hline
 \multirow{2}{*}{$15^{th}$} & Single & Fault-free & Full Coverage & 99.98\% & 99.76\% & 99.65\% \\   & Full & Fault-free & Full Coverage & Full Coverage & Full Coverage & Full Coverage \\ \hline
\end{tabular}
}
\vspace*{-1.5em}
\end{table}

\SetKwInOut{KwInOut}{In/Out}
\SetKwInOut{KwIn}{In}
\SetKwInOut{KwOut}{Out}
\begin{algorithm}[ht!]
\caption{Adaptive-ABFT strategy}
\label{alg-abft}
\SetKwFunction{FMain}{\ABFTOC}
\SetKwProg{Fn}{Function}{:}{\KwRet{$F^{GPU}_{desired}$, $SingleABFTCheck$, $FullABFTCheck$}}
\Fn{\FMain{}}{
\KwIn{Desired ABFT fault coverage $FC_{desired}$}
\KwIn{Desired GPU clock freq. $F^{GPU}_{desired}$}
\KwIn{Default GPU clock freq. $F^{GPU}_{BASE}$}
\KwIn{Predicted operation execution time $T'^{GPU}$}
$SingleABFTCheck \leftarrow FALSE$\\
$FullABFTCheck \leftarrow FALSE$\\
\While{($\lambda_{F^{GPU}_{desired}, 0D} > 0$ || $\lambda_{F^{GPU}_{desired}, 1D} > 0$ || $\lambda_{F^{GPU}_{desired}, 2D} > 0$)
$\&\&$ $!SingleABFTCheck$ $\&\&$ $!FullABFTCheck$} {
$T^{GPU}_{projected} = T'^{GPU} \times \frac{F^{GPU}_{desired}}{F^{GPU}_{BASE}}$ \\
\uIf{$FC_{single}(F^{GPU}_{desired}, T^{GPU}_{projected}) \geqslant  FC_{desired}$}{
    $SingleABFTCheck = TRUE$\\
  }
  \uElseIf{$FC_{full}(F^{GPU}_{desired}, T^{GPU}_{projected}) \geqslant  FC_{desired}$}{
    $FullABFTCheck = TRUE$\\
  }
  \Else{
    $F^{GPU}_{desired} = F^{GPU}_{desired} - 100MHz$\\
  }
}
}
\end{algorithm}

ABFT is based on the idea that if we encode a certain amount of matrix information in checksums before a matrix operation and apply the same matrix operation to checksums, the checksum relation would still hold for the resulting matrix. 
By verifying the checksum relations after the operation, we can detect and correct errors in the result matrix. 
Depending on how much information is encoded in checksums, the fault tolerance strength is different. 
As shown in \textbf{Figure~\ref{single-full-abft}}, there are two commonly schemes for checksum encoding: \Circled{1} Single side checksum encodes matrices along either rows or columns.
Since it only encodes the matrix in one dimension, it brings relative lower overhead. However, it can only efficiently tolerate 0D error pattern.
\Circled{2} Full checksum encodes matrices along both rows and column at the same time. Since it encodes matrices in both dimensions, it brings stronger protection i.e., both 0D and 1D error patterns. 
However, it also brings higher fault tolerance overhead. 

Given that the fault tolerance strength is limited, we must determine suitable ABFT protection according to the error rate and limit the clock frequency range to ensure all errors can be detected and corrected with a high probability. 
Otherwise, undetected or uncorrected errors would cause serious error propagation later, which requires recovery with high overhead. 
In this work, we find that it is useful to estimate the probability that a certain kind of ABFT can detect and correct all errors given different error rates at different overclocking frequencies. 
In order to do that, we first define an error rate function $R$ given clock frequency derived from our profiling results in~\textbf{Figure~\ref{profile-overclock}}:
$\lambda_{f, ErrType} = R(f, ErrType)$
where $\lambda$ is the error rate of a certain error type ($ErrType$). The error type can be 0D, 1D, or 2D. 
$f$ is the processor clock frequency. 
Assuming the rate is constant for a given clock frequency, we treat the distribution of probability errors that occur during a period of time as the Poisson distribution. 
So, the probability of having $k$ errors in a certain type during a period of time $T$ can be estimated using the Poisson distribution function:
$p=\frac{e^{-\lambda_{f, ErrType}T} (\lambda_{f, ErrType} T)^{i}}{i!}$.
Both single-side and full checksum encode the matrix for each matrix block individually.
They cannot tolerate more than one fault strike to a matrix block during one error detection interval (i.e., one iteration of matrix decomposition).
Assuming the matrix is of size $n$ with matrix block size $b$, single-side checksum ABFT can tolerate up to $S = \frac{n}{b}\times \frac{n}{b}$ 0D errors, as long as two 0D errors do not strike the same matrix block within one iteration of matrix decomposition.
Full checksum ABFT can tolerate up to $S$ 0D and 1D errors combined, as long as two 0D/1D errors do not strike the same matrix block within one iteration of matrix decomposition.
Assuming error occurs randomly and uniformly in time and space, we provide the theoretical estimation on the probability that ABFT can detect and correct all errors in one detection interval (i.e. \textit{Fault Coverage (FC)}).
\small
\begin{gather*}
FC_{single}(f, T) = \\ \left (\sum_{k=0}^{S}
\frac{e^{-\lambda_{f, 0D}T} (\lambda_{f, 0D} T)^{k}}{k!}\prod_{i=0}^{k}\frac{S-i}{S}\right ) e^{-\lambda_{f, 1D}T} e^{-\lambda_{f, 2D}T}
\\
FC_{full}(f, T) = \\ \left (\sum_{k=0}^{S}\sum_{j=0}^{S-k}
\frac{e^{-\lambda_{f, 0D}T} (\lambda_{f, 0D} T)^{k}}{k!}\frac{e^{-\lambda_{f, 1D}T} (\lambda_{f, 1D} T)^{j}}{j!}\prod_{i=0}^{k+j}\frac{S-i}{S}\right ) e^{-\lambda_{f, 2D}T}
\end{gather*}

\normalsize
\textbf{Table~\ref{clock-abft}} show the example estimation results based on different GPU overclocking frequencies and the execution time of the TMU operation in three selected iterations of the LU decomposition.
We define $FC > 99.9999\%$ as \textit{Full Coverage}.
Having the capability of fault coverage estimation, we propose an adaptive-ABFT scheme.
Unlike existing ABFT works, which enable ABFT during the entire matrix decomposition process, our adaptive-ABFT only enables ABFT error detection and correction when the error rate is above 0.
\textbf{Algorithm~\ref{alg-abft}} shows the adaptive-ABFT strategy.
We first check the error rate function in Line 4. 
If the rate of any kind of error is above zero, we check if applying ABFT can provide enough fault coverage (Line 5 - 9).
We prioritize single-side ABFT over full ABFT to lower fault tolerance overhead.
If none of the ABFT schemes can provide enough fault coverage, we progressively lower the GPU clock frequency (Line 11) until enough fault coverage is provided. 
Finally, we return the adjusted clock frequency together with flags indicating if we need to do a single or full ABFT check.
Please note \ABFTOC~would also work for CPU. We exclusively apply it to GPU in our algorithm since SDCs only occur to the GPU on our test system.

\subsection{Bi-directional slack reclamation (\BSR)}
\label{bi}

\begin{figure}[!ht]
    \centering
    \includegraphics[width=0.45\textwidth]{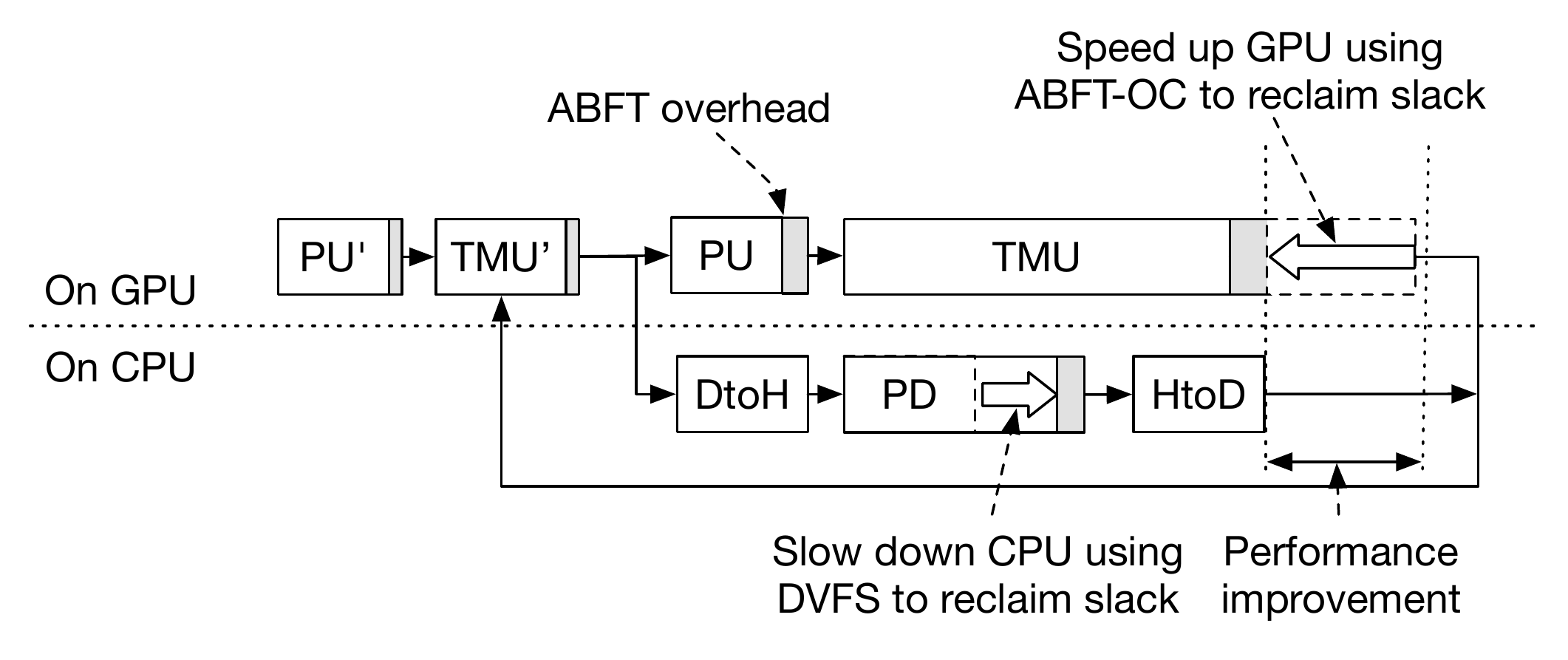}
    \vspace*{-1em}
    \caption{Bi-directional slack reclamation (\BSR)}
    \label{bi-slack}
    \vspace*{-1.5em}
\end{figure}

The current best energy-saving approach, single directional slack reclamation (\SR)~\cite{chen2016greenla}, saves energy by slowing down tasks on the non-critical paths via DVFS.
This work proposes a novel Bi-directional slack reclamation (\BSR) energy-saving technique that reclaims slacks in two directions at the same time using both \ABFTOC~and DVFS.
Specifically, \BSR~reclaims slacks by simultaneously slowing down tasks on the non-critical path using DVFS and speeding up tasks on the critical path using \ABFTOC.
An illustration of \BSR~is shown in \textbf{Figure \ref{bi-slack}}.
Compared with \SR, \BSR~brings three major advantages: \Circled{1} potential higher energy saving through both DVFS and \ABFTOC~at the same time; \Circled{2} performance improvement in addition to energy saving optimization; \Circled{3} enabling performance-energy consumption trade-off. 

\subsubsection{Enhanced Algorithmic-based Slack Prediction}
\label{slack-prediction}

\begin{table}[!ht]
\centering
\caption{Ratios of time complexity of PD, PU, TMU, transfer size, and ABFT-related operations between $k^{th}$ and $k+1^{th}$ iteration. $n$ and $b$ are the total size and the block size of the input matrix respectively. PU of Cholesky and QR are omitted since they do not affect the slack}
\label{ratios}

\label{inter-relation}
\resizebox{0.5\textwidth}{!}{%
\begin{tabular}{|c|p{2.5cm}|c|p{2.5cm}|}
\hline
Operation         & Computation \& Checksum Update & Data Transfer &  Checksum Verification\\ \hline
PD-Cho. & $1$ & $1$ & $1$ \\ \hline
TMU-Cho. & $(1+k)(1-\frac{b}{n-kb-b})$ & N/A  & $1-\frac{b}{n-kb-b}$\\ \hline
PD-LU       & $1-\frac{6b}{3n-(3k-1)b}$   &     $1-\frac{1}{n-kb}$  &     $1-\frac{1}{n-kb}$   \\ \hline
PU-LU       & $1-\frac{b}{n-kb-b}$   &     N/A  &     $1-\frac{b}{n-kb-b}$   \\ \hline
TMU-LU       &  $1-\frac{2b}{n-kb}$ & N/A  &    $1-\frac{2b}{n-kb}$          \\ \hline
PD-QR       & $1-\frac{b}{6n-(6k+1)b}$   &  $1-\frac{b}{n-kb-b}$     &      $1-\frac{b}{n-kb-b}$       \\ \hline
TMU-QR       & $1-\frac{b}{n-kb-b}-\frac{b}{n-kb+b}+\frac{b^2}{(n-kb-b)(n-kb+b)}$ &  N/A     &    $1-\frac{b}{n-kb-b}-\frac{b}{n-kb+b}+\frac{b^2}{(n-kb-b)(n-kb+b)}$         \\ \hline
\end{tabular}
}
\vspace*{-1em}
\end{table}
Slack prediction is critical for making correct power status adjustments so that energy saving can be maximized.
As \BSR~enables more opportunities for slack reclamation, it is more critical for it to make accurate slack predictions.
The state-of-the-art algorithmic slack prediction was first proposed by~\cite{chen2016greenla}. 
It mainly works by profiling the tasks in the $1^{st}$ iteration of decomposition and using the profiled time together with ratios of computational time complexity between $k^{th}$ iteration and the $1^{st}$ to predict the execution time of tasks in the $k^{th}$ iteration of decomposition.
By leveraging algorithmic knowledge and profiling results, algorithmic slack prediction can achieve much higher prediction accuracy compared with statistical-learning-based approaches and hardware-based approaches.

However, we find that the accuracy of current algorithmic slack prediction highly relies on the profiling accuracy of the $1^{st}$ iteration and the assumption that computational efficiency stays constant across different iterations on a given processor.
As the measurement of the $1^{st}$ iteration can be inaccurate (e.g., when it is short) and the computational efficiency of tasks can also change considerably throughout the decomposition process, all these inaccuracies can accumulate and cause large prediction errors in the latter part of the decomposition process, which lead to wrong slack reclamation decisions.

In \BSR, we propose an enhanced algorithmic-based slack prediction that greatly improves slack prediction accuracy.
The enhanced algorithmic-based slack prediction rely on the profiled execution time of the $p$ last neighbor iterations to predict the execution time of the current iteration to reduce the negative impacts bring by inaccurate profiling and changes in computational efficiency since tasks in neighbor iterations tend to have similar input sizes and thus similar computational efficiencies.
Since a closer neighbor has a more accurate estimation of computational efficiency, we apply different weights to different profiling results in our enhanced algorithmic-based slack prediction.
Specifically, the execution time of a task in $k^{th}$ iteration  ($T_k^{'OP}$) is predicted as:

$$T_k^{'OP} = w_1r^{OP}_{k-1,k}T_{k-1}^{OP}+w_2r^{OP}_{k-2,k}T_{k-21}^{OP}+...+w_pr^{OP}_{k-p,k}T_{k-p}^{OP}$$

where $r^{OP}_{j,k}$ is the ratio of theoretical time complexity of $OP$ between $j^{th}$ and $k^{th}$ iteration, which can be calculated based on the algorithm time complexity and relative change of the input sizes of $OP$.
\textbf{Table \ref{ratios}} shows the ratios of key components of matrix decompositions.
We omit the calculation process due to the page limit.
$T_{k-i}^{OP}$ is the actual profiled execution time of $OP$ of the $i^{th}$ last neighbor. 
$w_1$ is the weight we applied to the $i^{th}$ last neighbor.
Through empirical study, we find that $p=4$ and $w_1=\frac{1}{2}, w_2=\frac{1}{4},w_3=\frac{1}{8}, w_4=\frac{1}{8}$ can help provide enough prediction accuracy for energy saving. 
When ABFT is applied, the slack of the $k^{th}$ iteration is predicted as:
\vspace{-0.5em}
\begin{multline*}
slack_k = T_k^{'TMU} + T_k^{'TMU\ checksum\ update} +T_k^{'TMU\ checksum \ verf} \\
T_k^{'PU} + T_k^{'PU\ checksum\ update} +T_k^{'PU\ checksum \ verf} \\
- T_k^{'PD} - T_k^{'PD\ checksum\ update} - T_k^{'PD\ checksum\ verf} \\ - T_k^{'Data\ Transfer} - T_k^{'Transfer\ checksum}
\end{multline*}
\vspace{-1em}

\SetKwInOut{KwInOut}{In/Out}
\SetKwInOut{KwIn}{In}
\SetKwInOut{KwOut}{Out}
\begin{algorithm}[ht!]
\caption{\BSR~strategy}
\label{alg-bsr}
\SetKwFunction{FMain}{\BSR}
\SetKwProg{Fn}{Function}{:}{\KwRet{$AdjustCPU$, $AdjustGPU$, $F^{CPU}_{desired}$ ,$F^{GPU}_{desired}$, $SingleABFTCheck$, $FullABFTCheck$}}
\Fn{\FMain{}}{
\KwIn{reclamation ratio $r$}
\KwIn{iteration $k$}
\KwIn{GPU DVFS latency $L^{GPU}$}
\KwIn{CPU DVFS latency $L^{CPU}$}
\KwIn{Desired ABFT fault coverage $FC_{desired}$}
Apply optimized guardband for both CPU and GPU \\
$T'^{CPU}$,$T'^{GPU}$, $T'^{DataTransfer}$ $\leftarrow \textbf{\texttt{EnhancedAlgorithmicPrediction}}(k)$\\
$slack_k \leftarrow = T'^{GPU} - T'^{CPU} - T'^{DataTransfer}$ \\
\eIf{$slack_k > 0$}{
$T^{GPU}_{desired} \leftarrow   T'^{GPU} - (slack_k\times r) - L^{GPU}$ \\
$T^{CPU}_{desired} \leftarrow  T^{GPU}_{desired} - L^{CPU} - T'^{DataTransfer}$\\ 
}{
$T^{CPU}_{desired} \leftarrow   T'^{CPU} - (slack_k\times r) - L^{CPU}$ \\
$T^{GPU}_{desired} \leftarrow  T^{CPU}_{desired} - L^{GPU} + T'^{DataTransfer}$\\ 
}

$F^{GPU}_{desired} \leftarrow Roundup(F^{GPU}_{BASE} \times \frac{T'^{GPU}}{ T^{GPU}_{desired}}, 100Mhz)$\\
$F^{CPU}_{desired} \leftarrow Roundup(F^{CPU}_{BASE} \times \frac{T'^{CPU}}{ T^{CPU}_{desired}}, 100Mhz)$ \\
$F^{GPU}_{desired} = LimitToRange(F^{GPU}_{min}, F^{GPU}_{max})$ \\
$F^{CPU}_{desired} = LimitToRange(F^{CPU}_{min}, F^{CPU}_{max})$ \\
$T^{GPU}_{projected} = T'^{GPU} \times \frac{F^{GPU}_{desired}}{F^{GPU}_{BASE}}$ \\
$T^{CPU}_{projected} = T'^{CPU} \times \frac{F^{CPU}_{desired}}{F^{CPU}_{BASE}}$ \\
$T_{max} = max(T'^{GPU}, T'^{CPU} + T'^{DataTransfer}$)\\
\lIf{$T^{GPU}_{projected} > T_{max}$}{$AdjustGPU \leftarrow FALSE$}
\lElse{$AdjustGPU \leftarrow TRUE$}
\lIf{$T^{CPU}_{projected} > T_{max}$}{$AdjustCPU \leftarrow FALSE$}
\lElse{$AdjustCPU \leftarrow TRUE$}
$F^{GPU}_{desired}$, $SingleABFTCheck$, $FullABFTCheck$ $\leftarrow$ \textbf{\ABFTOC}($FC_{desired}$, $F^{GPU}_{desired}$, $F^{GPU}_{BASE}$, $T'^{GPU}$)
}
\end{algorithm}

\subsubsection{Bi-directional slack reclamation strategies}
\label{reclamation}
Compared with \SR, \BSR~offers more flexibility by reclaiming slacks from both directions, so the fractions of slacks that are reclaimed by the two tasks are adjustable, which in turn controls the performance-energy efficiency trade-off.
So, we define \textit{reclamation ratio} ($r$) to be the fraction of the slack we try to reclaim by speeding up the task on the critical path and $1-r$ to be the fraction we try to reclaim by slowing down the task on the non-critical path.
\textbf{Algorithm~\ref{alg-bsr}} shows our \BSR~algorithm that makes decisions at the beginning of each matrix decomposition iteration.
The execution time of tasks and slack are predicted in Line 3 - 4 using our enhanced algorithmic-based slack prediction.
Given reclamation ratio $r$, we calculate the desired execution time of tasks on CPU and GPU in Line 5 - 11.
We also consider the overhead of DVFS operations in our calculation to minimize the impact on performance.
Line 12 - 15 calculate the desired CPU/GPU clock frequencies and limit them within the available frequency range. 
Line 16 - 17 calculates the projected execution time if we apply the desired frequencies.
Note that the projected time may be different from the desired time since desired frequencies could be out of the available range.
Finally, we make decisions on whether or not we adjust CPU/GPU clock frequencies in Line 18 - 22.
If the projected time suggests that it can make a negative impact on the performance, it will skip frequency adjustment for this iteration i.e., setting AdjustCPU/GPU to FALSE.
Note that this does not mean we do not reclaim slack of this iteration. Since we still keep the adjusted CPU/GPU frequencies from the last iteration, the partial of slack can still be reclaimed.
This strategy ensures we reclaim most of the slacks while minimizing performance impact.
Line 23 invokes our adaptive-ABFT strategy for overclocking.
Finally, we return the final decisions regarding CPU/GPU clock frequency adjustments and ABFT protection strength for the current iteration.

\subsubsection{Theoretical performance improvement and energy saving analysis} 
Next, we provide a theoretical analysis of performance improvement and energy saving.
With losing generality, we assume that the slack on the CPU in the following discussion for simplification.
The performance improvement mainly comes from speeding up the tasks on the critical path.
So, the performance improvement of iteration $k$ can be simply calculated as: $\Delta T = T^{old}_k - T^{new}_k = T^{GPU}_k - (T^{GPU}_k - slack_k \times r) = slack_k \times r$. This suggests that higher $r$ leads to higher performance.
As for energy consumption, the theoretical amount of energy saving on the CPU when adopting \BSR~with reclamation ratio $r$ in the iteration $k$ can be estimated as:
\begin{gather*}
\Delta E^{CPU}_k = \Delta E^{CPU\_dynamic}_k +\Delta E^{CPU\_static}_k\\
\Delta E^{CPU\_dynamic}_k= E^{CPU\_dynamic\_old}_k - E^{CPU\_dynamic\_new}_k = \\
d^{CPU}P^{CPU}_{total}T_k^{CPU} - \\ \alpha^{CPU}\left(\frac{f^{CPU\_new}}{f^{CPU\_old}}\right)^{2.4}d^{CPU}P^{CPU}_{total}(T^{CPU}_k + slack_k(1-r)) =\\
d^{CPU}P^{CPU}_{total}T_k^{CPU} - \\ \alpha^{CPU}\left(\frac{T^{CPU}_k}{T^{CPU}_k + slack_k  (1-r)}\right)^{2.4}d^{CPU}P^{CPU}_{total}\\ \\ (T^{CPU}_k + slack_k (1-r))= \\
\left(1 - \alpha^{CPU}\frac{(T^{CPU}_k)^{1.4}}{(T^{CPU}_k + slack_k \times (1-r))^{1.4}}\right)d^{CPU}P^{CPU}T^{CPU}_k\\
\Delta E^{CPU\_static}_k=(T^{CPU}_k-\alpha^{CPU}(T^{CPU}_k + slack_k (1-r))) \\ (1-d^{CPU})P^{CPU}_{total}
\end{gather*}



Similarly, we can estimate the energy saving on GPUs as follows:
\begin{gather*}
\Delta E^{GPU}_k = \left (1-\alpha^{GPU}\frac{(T^{GPU}_k)^{1.4}}{(T^{GPU}_k-slack\times r)^{1.4}}\right )d^{GPU}P^{GPU}_{total}T^{GPU}_k + \\
(T^{GPU}_k-\alpha^{GPU}(T^{GPU}_k-slack_k\times r))(1-d^{GPU})P^{GPU}_{total}
\end{gather*}

Where $\alpha^{CPU/GPU}$ are total power reduction factors when we use optimized guardband of CPU/GPU.
We measure that in our hardware profiling work \textbf{Figure \ref{profile-overclock}}.
For clock frequencies out of the default range, we use constant values of the last measured value to estimate (dashed line).
$T^{CPU/GPU}_k$ are the original task execution time of CPU/GPU.
$P_{total}^{CPU/GPU}$ are the total power of CPU/GPU at the default guardband and clock frequencies.
$d^{CPU/GPU}$ are the ratios of the CPU/GPU dynamic power in the total power consumption.
The change of CPU/GPU dynamic power is estimated using: $P_{dynamic} \propto f^{2.4}$ \cite{cal14}.
When the critical path is on the GPU, it is for sure we can save energy on the CPU. 
However, whether or not we can save energy on the GPU depends on $\alpha^{GPU}$ and $r$.
Assuming power reduction factor $\alpha^{GPU}$ is fixed and minimized by applying optimized processor guardband, then the reclamation ratio $r$ controls the trade-off between performance improvement and energy consumption.
Higher $r$ leads to higher performance but less energy saving, and vice versa.
The highest energy saving can be achieved with $r_{max\_energy} = 0$ without performance improvement.
The max $r$ that achieves maximum without impacting energy efficiency is hard to be solved directly.
So, we use a numerical approach to solve for $r$. 
By solving $\Delta E^{CPU}_k + \Delta E^{GPU}_k = 0$ using Newton's method, we are able to get estimated solutions. For example, for decomposition with input $30730\times 30720$, the averaged reclamation ratios across all iterations are $0.28$ for Cholesky, $0.26$ for LU, and $0.31$ for QR, which approximately matches our experimental results in \textbf{Figure~\ref{perf-energy}}.


\section{Experimental Evaluation}
\label{experiments}
\subsection{Evaluation Methodology}
We compare~\BSR~with two state-of-the-art energy-saving approaches \RACE~and \SR~together with the original design in the MAGMA library.

\begin{itemize}
\item \texttt{Original}: The original matrix decompositions in the state-of-the-art MAGMA library. We keep the CPU/GPU clock frequency fixed at the default (autoboost disabled).
\item \RACE: The original matrix decompositions in the state-of-the-art MAGMA library with CPU/GPU autoboost feature enabled. The processor clock frequency is dynamically set according to the workload.
\item \SR: The state-of-the-art energy efficient matrix decompositions using single directional slack reclamation~\cite{chen2016greenla}.
\item \BSR: Our proposed matrix decomposition with \BSR~energy efficiency optimization and \ABFTOC. Clock frequencies can reach greater ranch where SDCs can occur but are correctable by ABFT.
\end{itemize}

All the above versions are implemented for Cholesky, LU, and QR decomposition for double precision inputs with block size tuned for performance.

\begin{table}[h!]
\scriptsize
\centering
\caption{Hardware/System Configuration for Experiments.}
\vspace{-1em}
\label{hardware_configuration}
\begin{tabular}{|p{1.5cm}|c|c|}
\hline
Processor & Intel Core i7-9700K & NVIDIA RTX 2080 Ti\\
\hline
Base Clock & 3.5($\uparrow$by0.1)GHz & 1.3($\uparrow$by 0.1)GHz\\
\hline
Overclocking & 3.6-4.5($\uparrow$by0.1)GHz & 1.4-2.2($\uparrow$by 0.1)GHz\\
\hline
Memory & 32 GB RAM & 12 GB RAM\\
\hline
Default guardband & Vcore offset: 0mV & Graphics clock offset: 0\\
\hline
Optimized guardband & Vcore offset: -150mV & Graphics clock offset: +200\\
\hline
\end{tabular}
\normalsize
\vspace{-1em}
\end{table}



\begin{figure}[ht]
    \centering
    \includegraphics[width=0.33\textwidth]{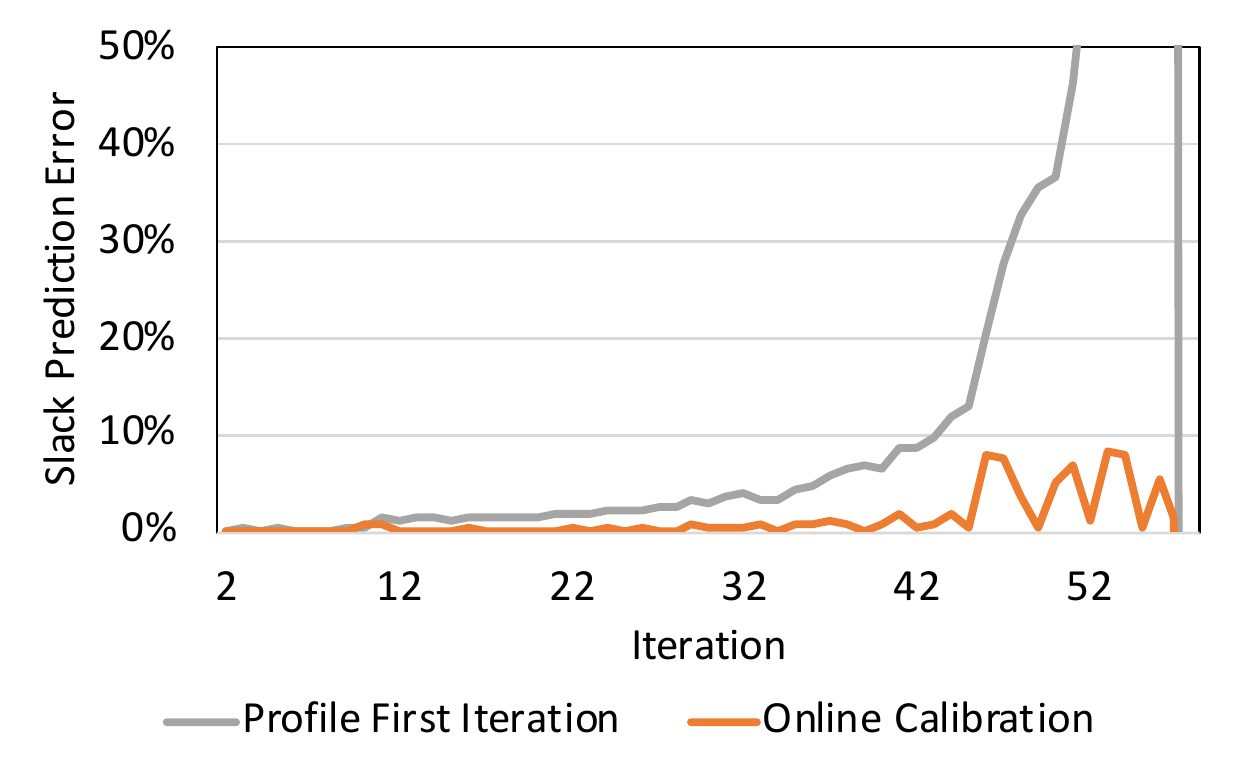}
    \vspace*{-1em}
    \caption{Slack prediction error of the LU decomposition using different approaches}
    \label{pred-error}
    \vspace*{-1em}
\end{figure}

\begin{figure}[ht]
    \centering
    \includegraphics[width=0.4\textwidth]{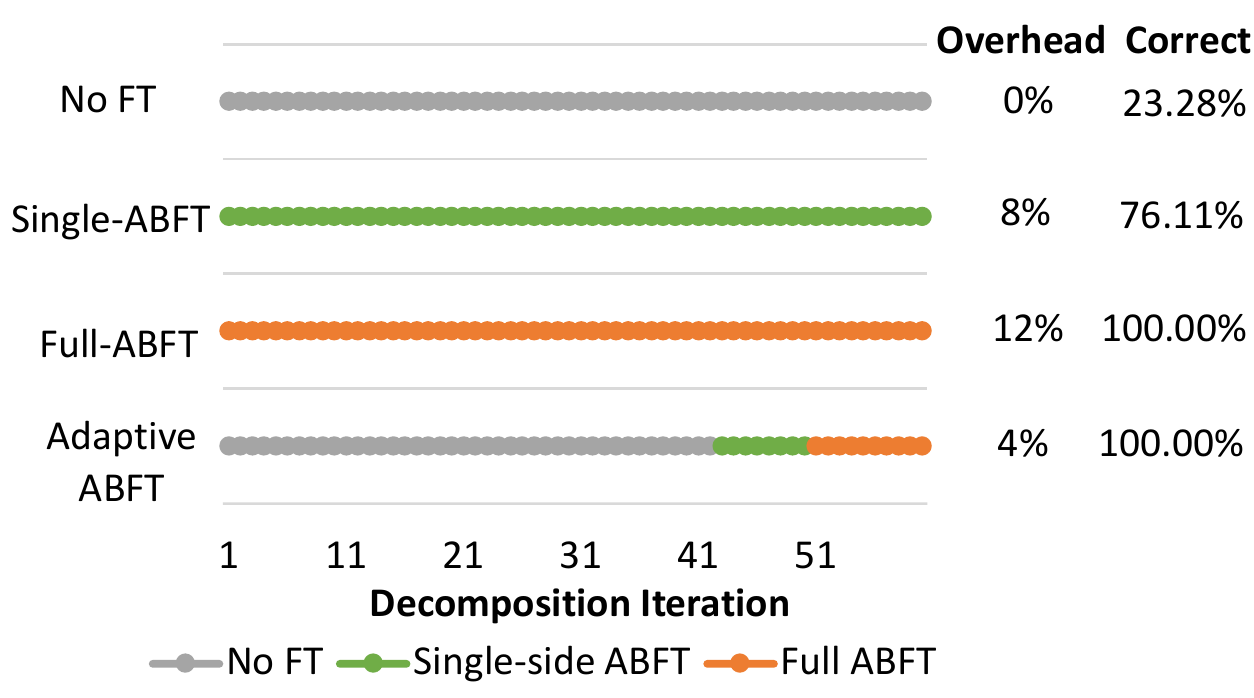}
    \vspace*{-1em}
    \caption{Comparing overhead and correctness when different ABFT scheme is applied in double precision LU decomposition with reclamation ratio $r = 0.25$}
    \label{abft}
    \vspace*{-1em}
\end{figure}

\begin{figure*}[h!]
    \captionsetup[subfigure]{aboveskip=-1pt,belowskip=-1pt}
    \centering
    \begin{subfigure}[t]{0.49\textwidth}
    \includegraphics[width=\textwidth]{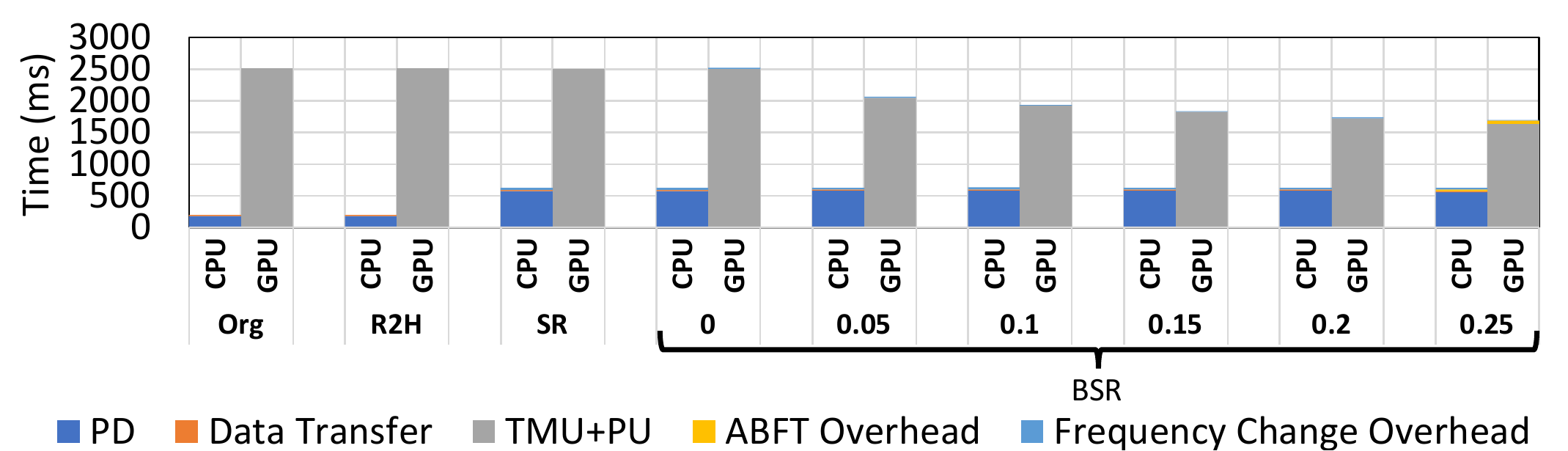}
    \caption{Time breakdown ($2^{nd}$ iteration)}
    \end{subfigure}
    \begin{subfigure}[t]{0.49\textwidth}
    \includegraphics[width=\textwidth]{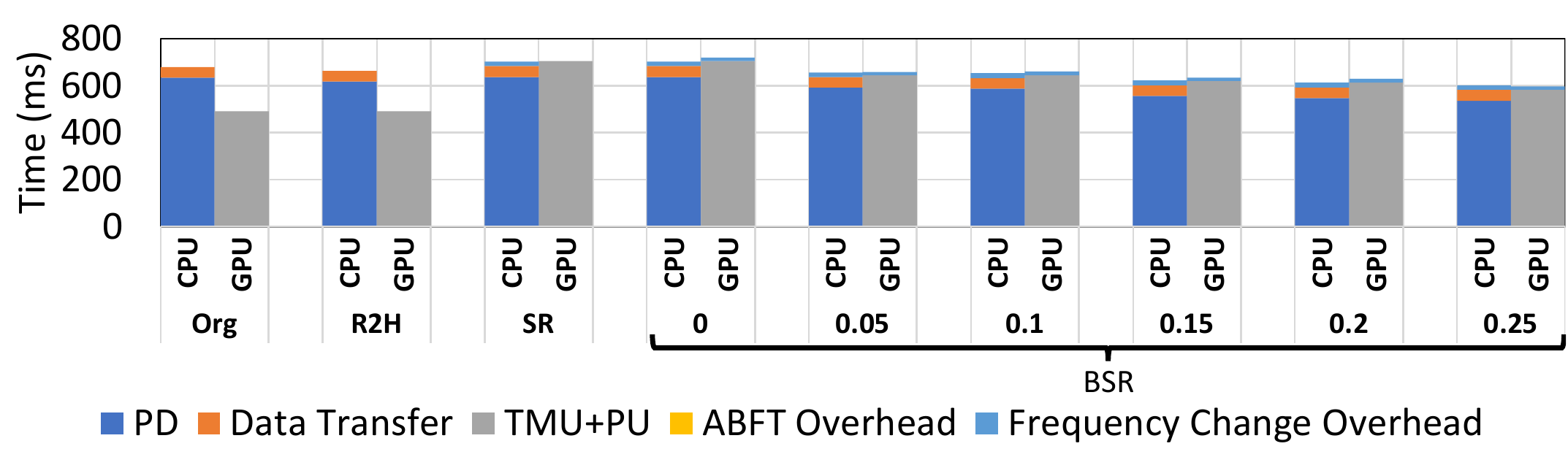}
    \caption{Time breakdown ($50^{th}$ iteration)}
    \end{subfigure}
     \begin{subfigure}[t]{0.49\textwidth}
    \includegraphics[width=\textwidth]{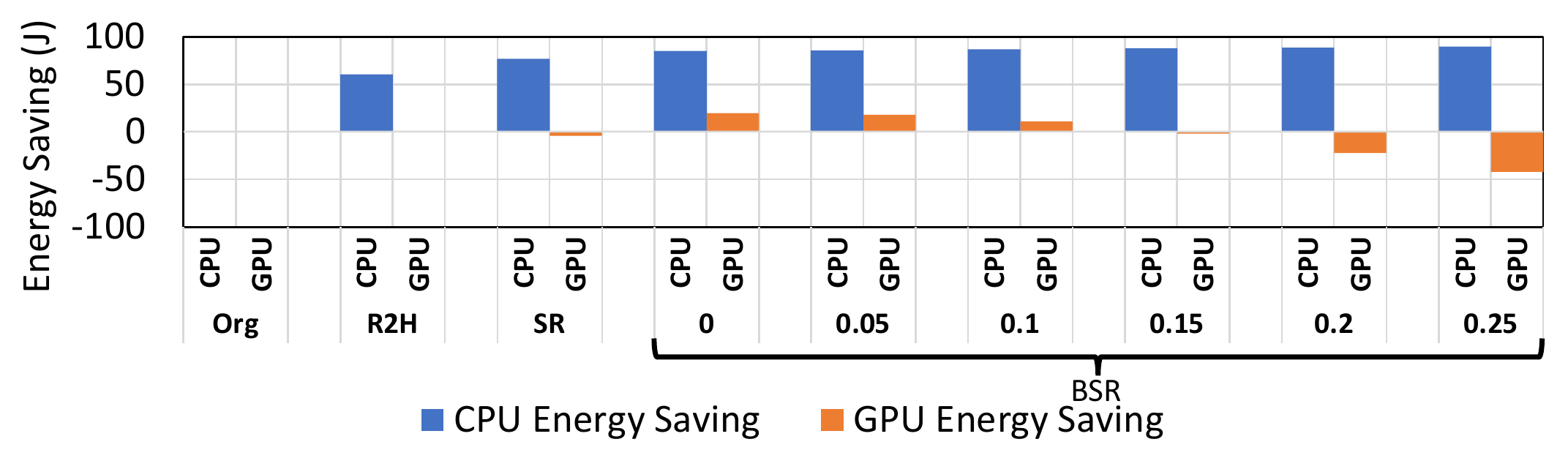}
    \caption{Energy saving breakdown ($2^{nd}$ iteration)}
    \end{subfigure}
    \begin{subfigure}[t]{0.49\textwidth}
    \includegraphics[width=\textwidth]{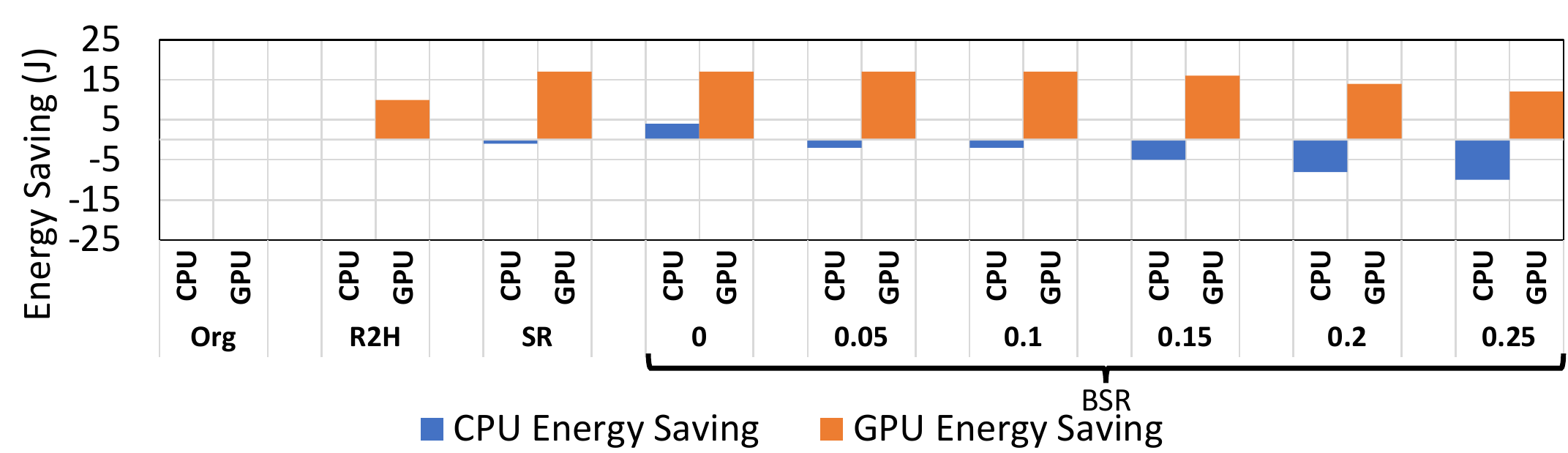}
    \caption{Energy saving breakdown ($50^{th}$ iteration)}
    \end{subfigure}
    \vspace*{-1em}
    \caption{Time and energy saving breakdown of the $2^{nd}$ and $50^{th}$ iteration of the LU decomposition (Input size: $30720 \times 30720$). Energy saving is compared with the original design. Positive values represent energy saving and negative values represent extra energy costs.}
    \label{1-iter}
    \vspace*{-1em}
\end{figure*}

\begin{figure*}[ht]
    \captionsetup[subfigure]{aboveskip=-1pt,belowskip=-1pt}
    \centering
    \begin{subfigure}[t]{0.33\textwidth}
    \includegraphics[width=\textwidth]{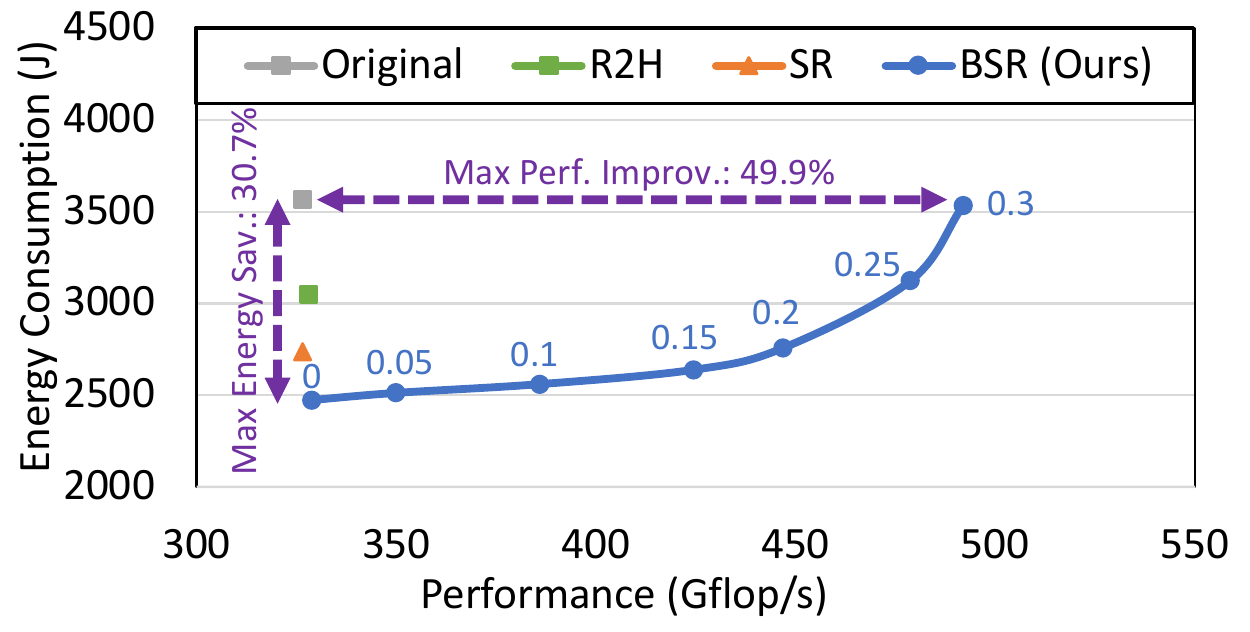}
    \caption{Cholesky}
    \end{subfigure}
    \begin{subfigure}[t]{0.33\textwidth}
    \includegraphics[width=\textwidth]{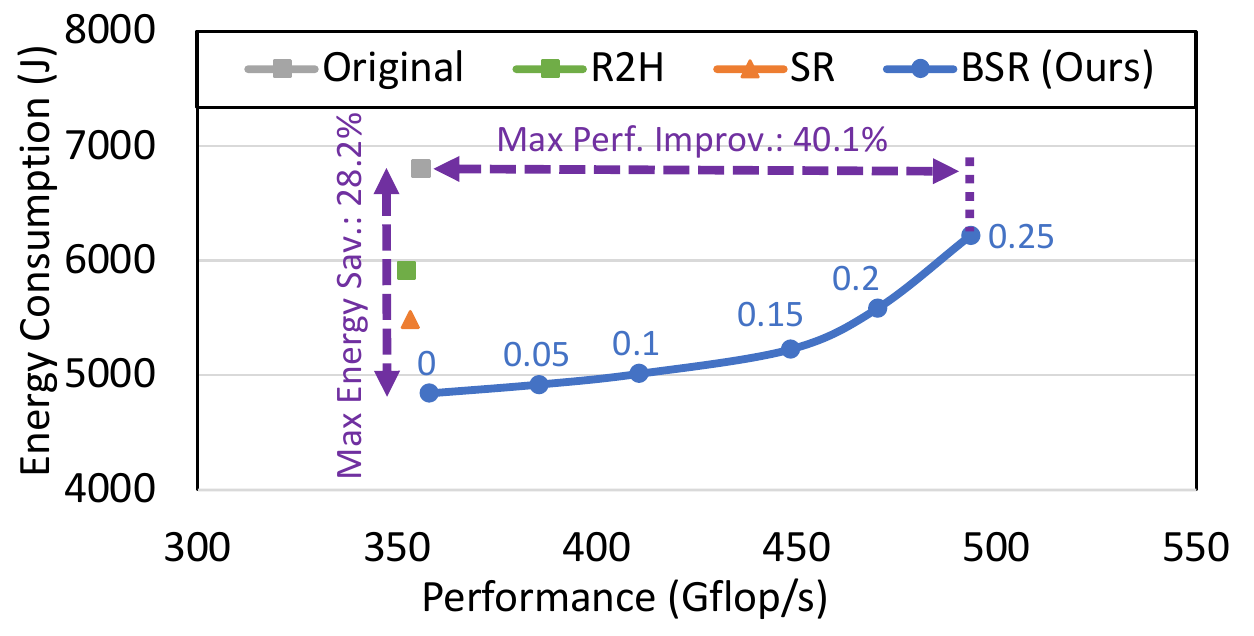}
    \vspace{-1em}
    \caption{LU}
    \end{subfigure}
    \begin{subfigure}[t]{0.33\textwidth}
    \includegraphics[width=\textwidth]{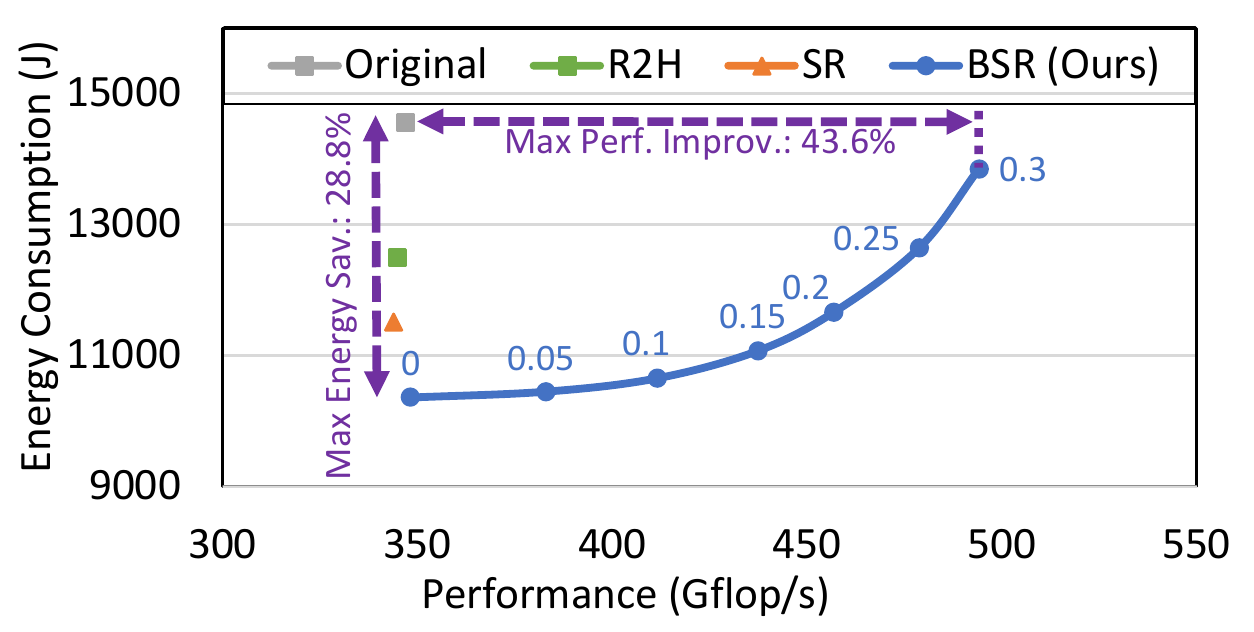}
    \vspace{-1em}
    \caption{QR}
    \end{subfigure}
    \vspace{-1em}
    \caption{Pareto efficient performance-energy consumption trade-off enabled by adjusting the reclamation ratio. Input size: $30720 \times 30720$ double precision }
    \label{perf-energy}
    \vspace*{-1em}
\end{figure*}

\begin{figure}[h!]
    \captionsetup[subfigure]{aboveskip=-1pt,belowskip=-1pt}
    \centering
    \begin{subfigure}[]{0.23\textwidth}
    \includegraphics[width=\textwidth]{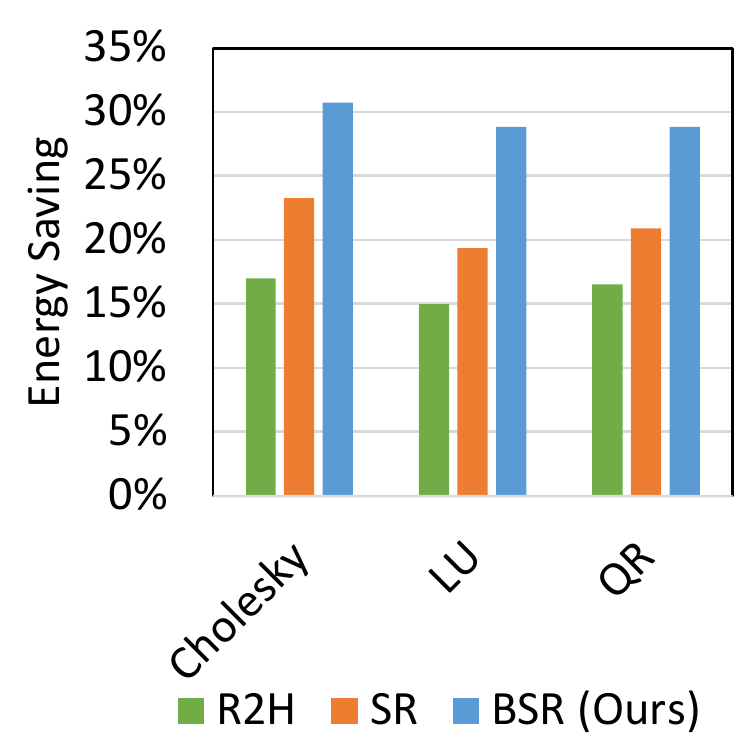}
    \caption{Energy Saving}
    \end{subfigure}
    \begin{subfigure}[]{0.23\textwidth}
    \includegraphics[width=\textwidth]{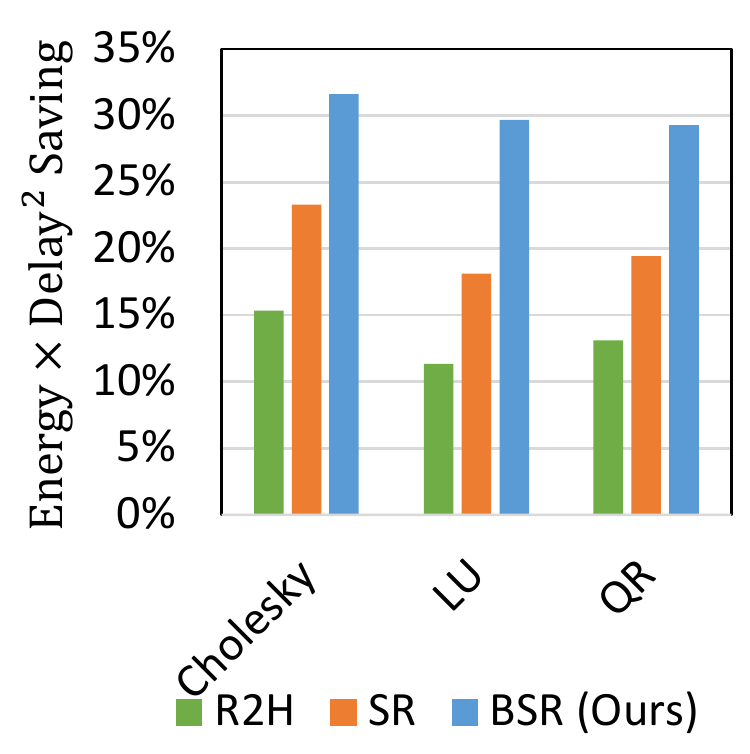}
    \caption{ED2P Reduction}
    \end{subfigure}
    \vspace*{-1em}
    \caption{Overall energy saving and ED2P Reduction compared with the original design. Input size: $30720 \times 30720$.}
    \label{overall-saving}
    \vspace*{-1em}
\end{figure}

\begin{figure}[t]
    \centering
    \includegraphics[width=0.4\textwidth]{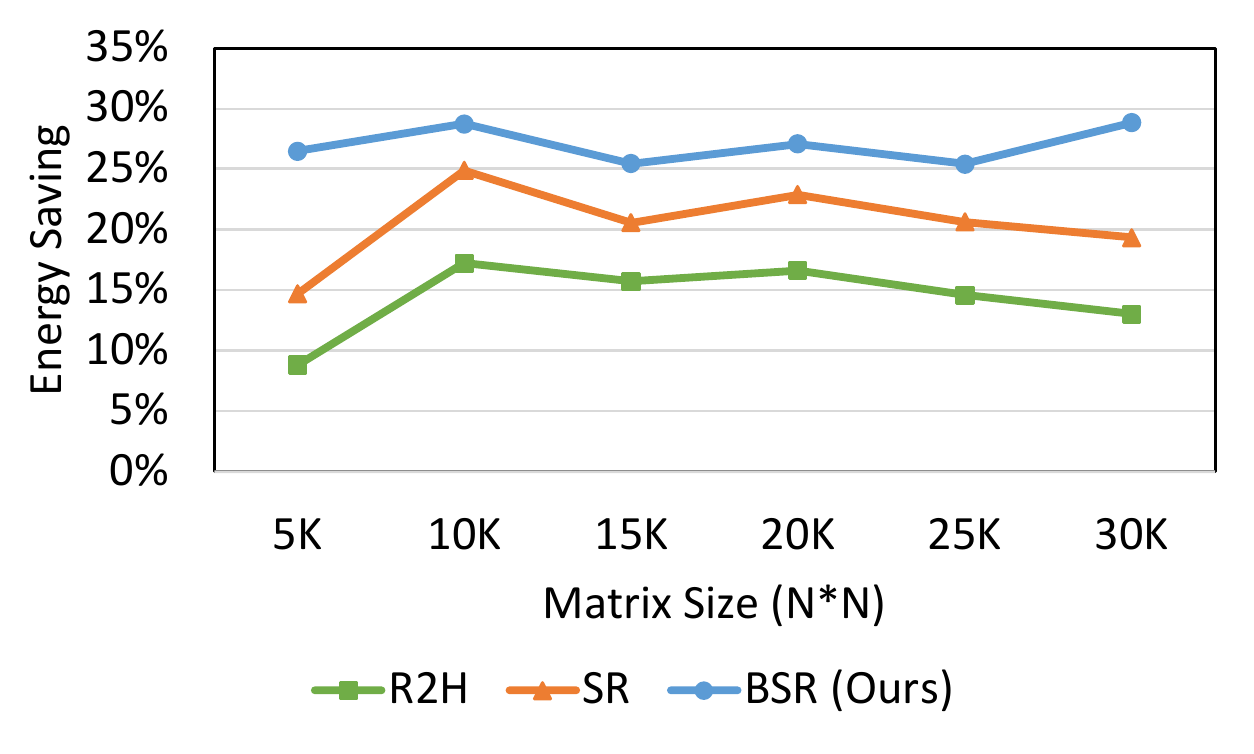}
    \vspace{-1em}
    \caption{Overall energy saving of LU compared with the original design with different input matrix sizes}
    \label{overall-size}
    \vspace{-2em}
\end{figure}

\subsection{Experimental Environment}
All experiments are performed on a power-aware CPU-GPU server. 
\textbf{Table~\ref{hardware_configuration}} lists hardware configuration of the experimental platform and system tools used for adjusting CPU/GPU guardband/clock frequencies and for measuring the energy consumption of CPU and GPU. 
Limited to the capability of our test platform, we only measure the energy consumption of the CPU package and GPU device. 
For accurate measurement of energy consumption and stable SDCs error rate at reduced guardband, we adjust the external cooling system to stabilize the CPU/GPU temperature at 45\textdegree{C} and 55\textdegree{C} respectively. 
From the software perspective, all matrix decomposition versions are built with GCC 7.4.0 and CUDA 11.6 with the highest optimization flags turned on.
NVIDIA cuBLAS 11.1 and Intel MKL 2020 are used as linear algebra computing kernels. 
MKL is configured to use all CPU cores. 
The operating system is Ubuntu 18.04.

\subsection{Evaluation Results}
\subsubsection{Online slack prediction accuracy comparison}
\textbf{Figure~\ref{pred-error}} shows the relative online prediction error using only the first iteration to predict~\cite{chen2016greenla} vs. our enhanced slack prediction approach proposed in this work. 
We can see both approaches can give less than 10\% relative error for the first 2/3 of the iterations. 
However, since~\cite{chen2016greenla} only depends on the profiling result of the first iteration, the error caused by profiling and prediction will accumulate and become significant (about 11.4\% on average) as the decomposition progresses. 
Our enhanced algorithmic slack prediction uses an online calibration approach to effectively avoid error from accumulating and reducing relative prediction error to around 4\% on average.

\subsubsection{ABFT overhead and correctness comparison}
\textbf{Figure~\ref{abft}} shows the computational overhead and probability of computing correctness when different ABFT schemes are applied.
We use double precision LU decomposition with \BSR~reclamation ratio $r = 0.25$ as an example.
The correctness is estimated by repeating the decomposition 100,000 times and comparing the results.
We observe similar results on other types of decompositions.
Due to relative short slack in the later part of decomposition, higher GPU clock frequencies are needed, which reach degrees of overclocking that can have SDC errors.
If we do not apply any fault tolerance, only 23.28\% of the overall matrix decomposition tests output correct results.
If we apply single-side checksum ABFT, it improves the percentage of tests with correct output to 76.11\% since 0D errors can be effectively detected and corrected.
However, 1D error cannot be handled by single-side checksum ABFT.
When full checksum ABFT is applied, it can ensure all decomposition tests are correct, but it also brings 12\% overhead.
Our adaptive-ABFT can adaptively apply necessary levels of fault tolerance to ensure high reliability and low overhead.
For example, when we set the reclamation ratio $r = 0.25$, the first 41 iterations are running at fault-free clock frequencies (1700Mhz), so adaptive-ABFT completely disables ABFT for eliminating unnecessary fault tolerance overhead.
For $42^{th}-49^{th}$ iteration, the slacks need to be reduced by \BSR~using more aggressive overclocking (up to 1900Mhz), so it applies single-side checksum ABFT.
Finally, it applies full checksum ABFT after $50^{th}$ iteration since higher clock frequencies are used (up to 2200Mhz).
So, with adaptive-ABFT, we can still ensure all decomposition tests are correct with only 4\% fault tolerance overhead.

\subsubsection{Per iteration performance and energy comparison}

To understand how each of the different approaches affects the performance and energy efficiency of matrix decompositions, we show the profiling results of $2^{nd}$ and $50^{th}$ iteration of the LU decomposition in terms of time and energy costs breakdown in \textbf{Figure~\ref{1-iter}}.
For the original version, we can see the slack occurs on the CPU side for the $2^{nd}$ iteration and GPU side for the $50^{th}$ iteration.
For clarity, we refer to the case that slack is on the CPU side as \Circled{C} and the case that slack is on the GPU side as \Circled{G} in our following discussion.
For \RACE, we observe noticeable energy saving in both \Circled{C} and \Circled{G} due to reduced energy consumption on the CPU side and GPU side respectively. 
For \SR, we see slack is fully reclaimed in \Circled{G}, but not fully reclaimed in \Circled{C} due to the limited clock frequency range on the CPU and longer slack length.
For \BSR, we test different reclamation ratios $r$ and mark their values under the bars.
We set $r$ from 0 to a certain value that leads to maximum achievable performance.
This maximum $r$ is higher for \Circled{C} than \Circled{G} since GPU has greater overclocking capabilities than CPU in our system when we apply optimized guardband. 
We can see maximum energy saving is achieved when $r = 0$, which is consistent with our previous theoretical analysis.
Maximum performance $r = 0.25$ for \Circled{C} and \Circled{G}, which are close to our theoretical estimation. 
When we increase $r$, we see an increase in energy consumption for the processor on the critical path due to the increase in clock frequency.
For \Circled{C}, we observe a slight increase in energy-saving since the slack is long enough for the CPU to always run at the lowest clock frequency, and reducing the total execution time can save more CPU static energy.
We also observe a slight decrease in energy saving in \Circled{G}, mainly due to the slight increases in clock frequencies.
Even though it can still save energy since 1) the clock frequencies are low; 2) power reduction brings by optimized guardband.
Finally, Thanks to \ABFTOC, we can exploit higher overclocking frequencies where we can achieve higher performance and energy efficiency in \Circled{C}.

\subsubsection{Overall energy saving and energy efficiency comparison}
Next, we show the overall energy-saving capability of different approaches in \textbf{Figure~\ref{overall-saving}(a)}.
We evaluate all three matrix decompositions with an input size of $30720\times 30720$.
All four versions of each type of matrix decomposition produce a similar performance. 
To maximize energy saving the reclamation ratio of \BSR~is set to 0.
We can see that compared with the state-of-the-art MAGMA library, our \BSR~is able to save energy by 30.7\% for Cholesky, 28.2\% for LU, and 28.8\% for QR.
That is $1.31\times - 1.49\times$ more energy saving compared with the current state-of-the-art \SR~energy saving approach and $2.03\times - 2.20\times$ more energy saving compared with \RACE.
In addition, we use $Energy \times Delay^2$ (ED2P) to measure the energy efficiency of matrix decompositions.
As shown in \textbf{Figure~\ref{overall-saving}(b)}, compared with the original design, our \BSR~is able to reduce ED2P by  29.3\%-31.6\%. 
Compared with \RACE, \BSR~is able to reduce ED2P by 18.6\%-20.7\%.
Finally, compared with \SR, \BSR~is able to reduce ED2P by 10.8\%-14.1\%.
\subsubsection{Overall energy saving on different input sizes}

In \textbf{Figure~\ref{overall-size}}, we show the results of applying energy-saving approaches on LU decomposition with different input sizes.
Limited by the page space, we only show the results for LU decomposition. Other matrix decompositions behave similarly.
We can see our \BSR~is able to stably save energy consumption across different input matrix sizes ranging from $5120\times5120$ and above.
Note that it is hard to save energy on smaller matrices since they either lead to high fault tolerance overhead or small slacks that are hard to be reclaimed.

\subsubsection{Overall Pareto efficient performance-energy consumption trade-off}

Finally, we show the overall Pareto efficient performance-energy consumption trade-off enabled by adjusting the reclamation ratio in \BSR.
As shown in \textbf{Figure~\ref{perf-energy}}, by adjusting the reclamation ratio to a minimum 0, we achieve max energy saving with similar performance to the original design.
In this case, compared with the original design, \BSR~is able to save energy by 28.2\%-30.7\%.
Compared with \RACE, \BSR~is able to save energy by 17.1\%-18.9\%.
Compared with \SR, \BSR~is able to save energy by 9.6\%-11.7\%.
By increasing the reclamation ratio, we are able to adjust the performance or energy consumption of matrix decompositions.
For example, with equal or less energy consumption, compared with the original design \BSR~is enable to improve the performance by 1.38$\times$-1.51$\times$.
Also, compared with \RACE, \BSR~is enable to improve the performance by 1.33$\times$-1.43$\times$.
In addition, compared with \SR, \BSR~is enable to improve the performance by 1.36$\times$-1.43$\times$.
Finally, we see the results of \BSR~with different reclamation ratios form a Pareto set such that we cannot improve energy saving and performance at the same time without reliability degradation.

\section{Conclusion}
\label{conclusion}
In this work, we focused on further improving the energy saving of matrix decompositions on CPU-GPU heterogeneous systems beyond existing state-of-the-art works.
To achieve our goal, we first proposed \ABFTOC, a novel overclocking technique that is protected by ABFT to enable reliable computation for key operations in matrix decompositions when overclocking. 
Next, based on \ABFTOC, we proposed \BSR, a novel matrix decomposition framework, that aims to maximize energy saving while maintaining performance and reliability.
We evaluated \BSR~on three key matrix decomposition algorithms - Cholesky, LU, and QR.
Experiments show that \BSR~is able to save up to 11.7\% more energy compared with the current best energy saving optimization approach with no performance degradation and up to 14.1\% ED2P reduction.
Also, \BSR~enables the Pareto efficient performance-energy trade-off, which is able to provide up to 1.43$\times$ performance improvement without costing extra energy.

\section{Acknowledgement}
The fault tolerance developments in this work by Z. C. were supported by the U.S. Department of Energy, Office of Science, Office of Advanced Scientific Computing Research, Scientific Discovery through the Advanced Computing (SciDAC) program under Award Number DE-SC0022209. The energy-saving techniques in this work by L. B. was supported by the National Science Foundation under Grant Number 1907401.

\bibliographystyle{ACM-Reference-Format}
\bibliography{biblio}
\end{document}